\documentclass[referee]{aa} 
\usepackage{amsmath}
\usepackage{amssymb}
\usepackage{graphicx}
\usepackage{txfonts}
\begin{document}

   \title{Magnetars: \\
       Structure and evolution from p-star models}
   \author{Paolo Cea
          \inst{1,2}
          }

   \offprints{Paolo Cea}

   \institute{Dipartimento Interateneo di Fisica, Universit\`a di Bari,
              Via G. Amendola 173, I-70126 Bari\\
              \email{Paolo.Cea@ba.infn.it}
         \and
             INFN - Sezione di Bari, Via G. Amendola 173, I-70126 Bari
             }

\date{Received / Accepted}

\abstract{ P-stars are compact stars made of up and down quarks in $\beta$-equilibrium with electrons in a chromomagnetic
condensate. We discuss p-stars endowed with super strong dipolar magnetic field which, following  consolidated tradition
in literature, are referred to as magnetars. We show that soft gamma-ray repeaters and anomalous $X$-ray pulsars can be
understood within our theory. We find a well defined criterion to distinguish rotation powered pulsars from magnetic
powered pulsars. We show that glitches, that in our magnetars are triggered by magnetic dissipative effects in the inner
core, explain both the quiescent emission and bursts in soft gamma-ray repeaters and anomalous $X$-ray pulsars. We account
for the braking glitch from SGR 1900+14 and the normal glitch from AXP 1E 2259+586 following a giant burst. We discuss and
explain the observed anti correlation between hardness ratio and intensity. Within our magnetar theory we are able to
account quantitatively for  light curves for both gamma-ray repeaters and anomalous $X$-ray pulsars. In particular we
explain the puzzling light curve after the June 18, 2002 giant burst from AXP 1E 2259+586.
}
\maketitle
%
%

\section{Introduction}
\label{Introduction}
In few years since their discovery (\cite{hewish:1968}), pulsars have been identified with rotating neutron stars, first
predicted theoretically by Baade \& Zwicky (1934a, 1934b, 1934c), endowed with a strong magnetic field
(\cite{pacini:1968,gold:1968}). Nowadays, no one doubts that pulsars are  neutron stars, even though it should be
remembered that there may be other alternative explanations for pulsars.  Up to present time it seems that there are no
alternative models able to provide as satisfactory an explanation for the wide variety of pulsar phenomena as those built
around the rotating neutron star model. However, quite recently we have proposed (\cite{cea:2003},b) a new class of
compact stars, named p-stars,  made of up and down quarks in $\beta$-equilibrium with electrons in
an abelian chromomagnetic condensate which is challenging the standard model based on neutron stars. \\
In the present paper we  investigate the properties of anomalous $X$-ray pulsars (AXPs) and soft gamma-ray repeaters
(SGRs). For a recent review on the observational properties of anomalous $X$-ray pulsars see Mereghetti (1999), Mereghetti
et al. (2002), Kaspi \& Gavriil (2004), for soft gamma-ray repeaters see Hurley (1999), Woods (2003). Recently, these two
groups have been linked by the discovery of persistent emission from soft gamma ray repeaters that is very similar to
anomalous $X$-ray pulsars, and bursting activity in anomalous $X$-ray pulsars quite similar to  soft gamma ray repeaters
(see, for instance \cite{Kaspi:2004b,Woods:2004}).
Duncan \& Thompson (1992) and Paczy\'nski (1992) have proposed that soft gamma-ray repeaters are pulsars whose surface
magnetic fields exceed the critical magnetic field $B_{QED} \simeq  4.4 \; 10^{13} \; \; Gauss$. Indeed, Duncan \&
Thompson (1992)  refer to these pulsar as {\it{magnetars}}. In particular Duncan \& Thompson (1995, 1996) argued that the
soft gamma-ray repeater bursts and quiescent emission were powered by the decay of an ultra-high magnetic field. This
interpretation is based on the observations that showed that these peculiar pulsars are slowing down rapidly, with an
inferred magnetic dipole field much greater than the quantum critical field $B_{QED}$, while producing steady emission at
a rate far in excess of the rotational kinetic energy loss. \\
The main purpose of this paper is to discuss in details p-stars endowed with super strong dipolar magnetic field which,
following well consolidated tradition in literature, will be referred to as magnetars. We will show that soft gamma-ray
repeaters and anomalous $X$-ray pulsars can be understood within our theory. The plan of the paper is as follows. In
Section~\ref{rotation} we discuss the phenomenological evidence for the dependence of pulsar magnetic fields on the
rotational period. We argue that there is a well defined criterion which allows us to distingue between rotation powered
pulsars and magnetic powered pulsars. We explicitly explain why the recently discovered high magnetic field radio pulsars
are indeed rotation powered pulsars. In Section~\ref{death}  we introduce the radio death line, which in the $\dot{P} - P$
plane separated radio pulsars from radio quiet magnetic powered pulsars, and compare with available observational data.
Section~\ref{glitches} is devoted to the glitch mechanism in our magnetars and their observational signatures. In
Section~\ref{braking} we compare glitches in SGR 1900+14 and 1E 2259+586, our prototypes for soft gamma-ray repeater and
anomalous $X$-ray pulsar respectively. Sections~\ref{quiescent} and \ref{bursts} are devoted to explain the origin of the
quiescent luminosity, the bursts activity and the emission spectrum during bursts. In Section~\ref{hardness} we discuss
the anti correlation between hardness ratio and intensity. In Section~\ref{light} we develop a general formalism to cope
with light curves for both giant and intermediate bursts. In Sections~\ref{2259} through \ref{1806} we careful compare our
theory with the available light curves in literature. In particular, we are able  to account for the peculiar light curve
following the June 18, 2002 giant burst from the anomalous $X$-ray pulsar 1E 2259+586. Finally, we draw our conclusions in
Section~\ref{conclusion}.
\section{\normalsize{ ROTATION VERSUS MAGNETIC POWERED PULSARS}}
\label{rotation}
\begin{figure}[t]
   \centering
\includegraphics[width=0.9\textwidth,clip]{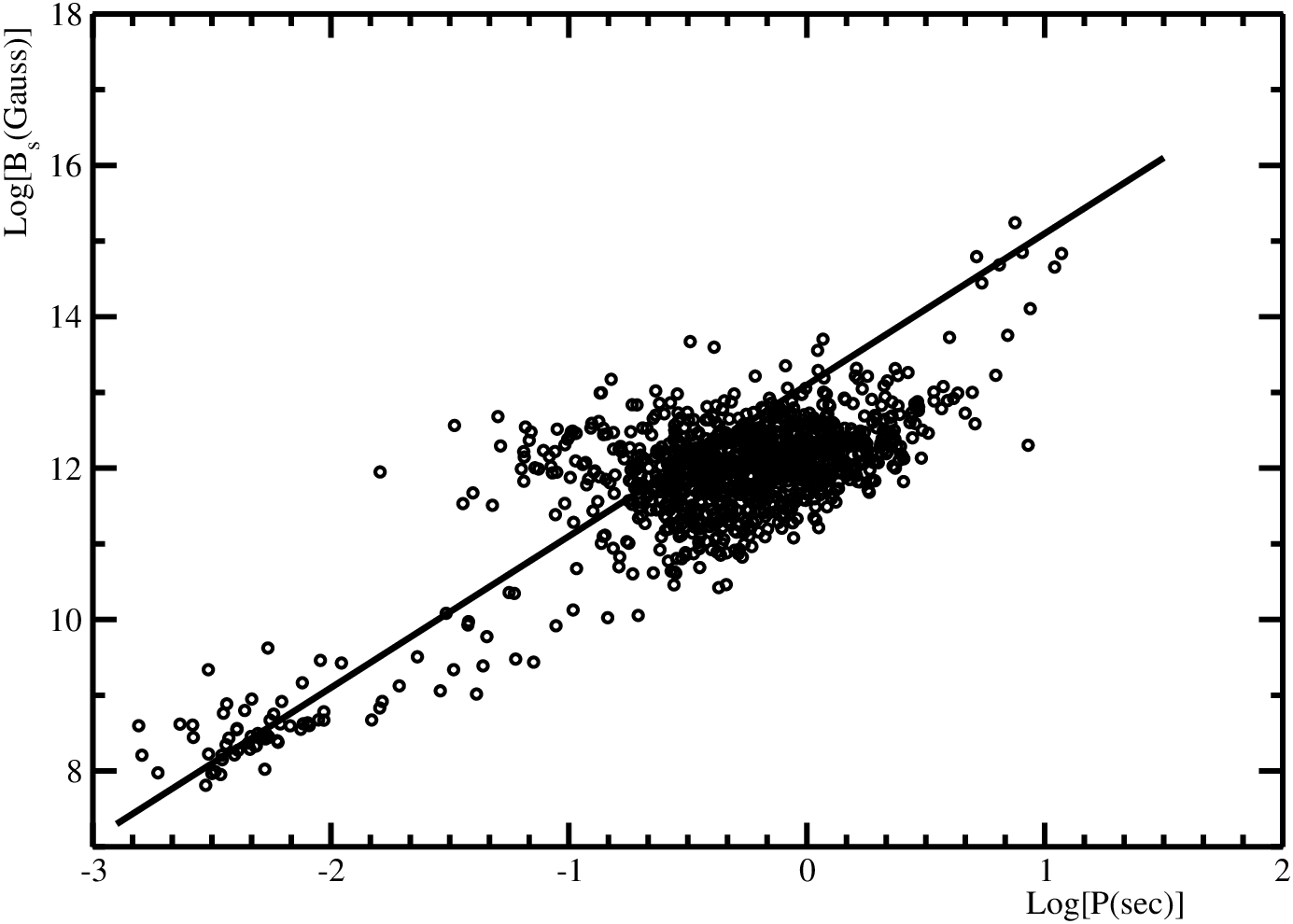}
\caption{\label{fig_1}
Inferred magnetic field $B_S$  plotted versus stellar period for 1194 pulsars taken from the ATNF Pulsar Catalog
(\cite{ATNF}). Full  line corresponds to Eq.~(\ref{magn-period}) with $B_1 \, \simeq \, 1.3 \, 10^{13} \; Gauss$.}
\end{figure}

As discussed in  Cea (2004a) the structure of p-stars is determined once the equation of state appropriate for the
description of deconfined quarks and gluons in a chromomagnetic condensate is specified.
In general, the quark chemical potentials are smaller that the strength of the chromomagnetic condensate. So that, up and
down quarks occupy the lowest Landau levels.  However, for certain values of the central energy  density it happens that a
fraction of down quarks must jump into higher Landau levels  in the stellar core, leading to a central core with energy
density $\varepsilon_c$ somewhat greater than the energy density outside the core. Now, these quarks in the inner core
produce a vector current in response to the chromomagnetic condensate. Obviously, the quark current tends to screen the
chromomagnetic condensate by a very tiny amount. However, since the down quark has an electric charge $q_d  =  -
\frac{1}{3} \, e$ ($e$ is the electric charge), the quark current generates in the core a uniform magnetic field parallel
to the chromomagnetic condensate with strength $B_c  \simeq  \frac{e}{96 \, \pi }  \;  gH$ (here and in the following we
shall adopt natural units $\hbar \; = \; c \; = \; k_B \; = \; 1$).
The inner core is characterized by huge conductivity, while outer core quarks are freezed into the lowest Landau levels.
So that, due to the energy gap between the lowest Landau levels and the higher ones, the quarks outside the core cannot
support any electrical current. As a consequence, the magnetic field in the region outside the core is dipolar leading to
the surface magnetic field:
\begin{equation}
\label{magn-core-surf}
B_S\; \; \simeq \; \;B_c \; \; \left (\frac{R_c}{R } \right )^3 \;  \; ,  \;  \; B_c \; \simeq \; \; \frac{e}{96 \, \pi }
\; \; gH\; \; ,
\end{equation}
$R$ and $R_c$ being the stellar and inner core radii respectively. In general the formation of the inner core denser than
the outer core is contrasted by the centrifugal force produced  by stellar rotation. Since the centrifugal force is
proportional to the square of the stellar rotation frequency, this leads us to argue that the surface magnetic field
strength is proportional to the square of the stellar period:
\begin{equation}
\label{magn-period}
B_S\; \; \simeq \; \; B_1 \; \left (\frac{P}{1 \, sec } \right )^2 \; \; \; \; ,
\end{equation}
where $B_1$ is the surface magnetic field for pulsars with nominal period $P \, = \, 1 \; sec$. Indeed, in
Fig.~\ref{fig_1} we we display the surface  magnetic field strength $B_S$ inferred from (for instance, see
\cite{manchester:1977}):
\begin{equation}
\label{magn-surf}
B_S\; \; \simeq \; \; 3.1 \; 10^{19} \; \; \sqrt{P \;\dot{P} }  \; Gauss \; \; ,
\end{equation}
versus the period. Remarkably, assuming $B_1 \, \simeq \, 1.3 \, 10^{13} \; Gauss$, we find the Eq.~(\ref{magn-period})
accounts rather well the inferred magnetic field for pulsars ranging from millisecond pulsars up to anomalous $X$-ray
pulsars and soft-gamma repeaters.
As a consequence of Eq.~(\ref{magn-period}), we see that the dipolar magnetic field is time dependent. In fact, it is easy
to find:
\begin{equation}
\label{mag-time}
 B_S(t) \; \simeq \;  B_0 \; \; \left ( 1 \; + \; 2 \; \frac{\dot{P}}{P} \; t \;  \right )\;
 \;  \; \; ,
\end{equation}
where $B_0$ indicates the magnetic field at the initial observation time. Note that Eq.~(\ref{mag-time}) implies that the
magnetic field varies on a time scale given by the characteristic age.  \\
It is widely accepted that pulsar radio emission is powered by the rotational energy:
\begin{equation}
\label{ener-rot}
E_{R} \; = \;  \frac{1}{2} \; I \; \; \omega^2 \; \; \; ,
\end{equation}
so that, the spin-down power output is given by:
\begin{equation}
\label{ener-rot-dot}
- \; \dot{E}_{R} \; = \; - \; I \; \; \omega \; \dot{\omega} \; \; = \; 4 \;
 \pi^2 \; I \; \frac{\dot{P}}{P^3} \; .
\end{equation}
On the other hand, an important source of energy is provided by the magnetic field. Indeed, the classical energy stored
into the magnetic field is:
\begin{equation}
\label{ener-mag}
 E_{B} \; = \; \frac{1}{8 \, \pi} \;  \int_{r \, \geq \, R} \; \; d^3r \; B^2(r) \; \;  \; \; ,
\end{equation}
Assuming a dipolar magnetic field:
\begin{equation}
\label{dip-mag}
 B(r) \; = \;  B_S \; \; \left ( \frac{R}{r} \right )^3 \; \; for \; \; \;
 r \, \geq \, R \; \;  \; \; ,
\end{equation}
Eq.~(\ref{ener-mag}) leads to:
\begin{equation}
\label{ener-mag-dip}
 E_{B} \; = \; \frac{1}{6 } \; B_S^2 \; \; R^3 \; \; .
\end{equation}
Now, from Eq.~(\ref{mag-time}) the surface magnetic field  is time dependent. So that, the magnetic power output is given
by:
\begin{equation}
\label{ener-mag-dot}
 \dot{E}_{B} \; = \; \frac{2}{3} \;  B_0^2 \; \; R^3 \; \;
  \; \frac{\dot{P}}{P} \; .
\end{equation}
For rotation-powered pulsars it turns out that $ \dot{E}_{B} \; \ll \; - \, \dot{E}_{R}$. However, if the dipolar magnetic
field is strong enough, then the magnetic power Eq.~(\ref{ener-mag-dot}) can be of the order, or even greater than the
spin-down power. Thus, we may formulate a well defined criterion to distinguish rotation  powered pulsars from  magnetic
powered pulsars. Indeed, until $ \dot{E}_{B} \; < \; - \, \dot{E}_{R}$ there is enough rotation power to sustain the
pulsar emission. On the other hand, when $ \dot{E}_{B} \; \geq \; - \, \dot{E}_{R}$ all the rotation energy is stored into
the increasing magnetic field and the pulsar emission is turned off. In fact, in the next Section we will derive the radio
death line, which is the line that in the $P-\dot{P}$ plane separates rotation-powered pulsars from magnetic-powered
pulsars. In the remainder of this Section, we discuss the recently detected radio pulsars with very strong surface
magnetic fields. We focus on the two radio pulsars with the strongest surface magnetic field: PSR J1718-3718 and PSR
J1847-0130. These pulsars have inferred surface magnetic fields well above the quantum critical field $B_{QED}$ above
which some models (\cite{Baring:1998}) predict that radio emission should not occur. In particular, we have:
\begin{equation}
\label{1718-1847}
\begin{split}
PSR J1718-3718 \; \; \; (\cite{Hobbs:2004}) &  \; \; \; \; \; \; \; P \; \simeq \; \; \; 3.4 \; \; sec  \; \; , \; \;
      B_S \, \simeq \, 7.4 \; 10^{13} \; Gauss \; \; , \\
PSR J1847-0130 \; \; (\cite{McLaughlin:2003})  &  \; \; \; \; \; \; \; P \; \simeq \; \; \; 6.7 \; \;  sec  \; \; , \; \;
    B_S \, \simeq \, 9.4 \; 10^{13} \; Gauss \; \; .
\end{split}
\end{equation}
Both pulsars have average radio luminosities and surface magnetic fields larger than that of AXP 1E 2259+586. Now, using
Eqs.~(\ref{ener-mag-dot}), (\ref{ener-rot-dot}) , together with $I = \frac{2}{5} \, M \, R^2$, and Eq.~(\ref{1718-1847})
we get:
\begin{equation}
\label{1718-1847-power}
\begin{split}
  PSR J1718-3718 &  \; \; \; \;  - \; \dot{E}_{R} \;  \simeq \; 3.4 \; 10^{45} \;  erg \; \;  \;
  \frac{\dot{P}}{P} \; , \; \dot{E}_{B} \;  \simeq \; 4.7 \; 10^{44} \;  erg \; \;  \;
  \frac{\dot{P}}{P} \;  \; , \\
  PSR J1847-0130  &  \; \; \; \;  - \; \dot{E}_{R} \;  \simeq \; 8.8 \; 10^{44} \;  erg \; \;  \;
  \frac{\dot{P}}{P} \; , \; \dot{E}_{B} \;  \simeq \; 7.5 \; 10^{44} \;  erg \; \;  \;
  \frac{\dot{P}}{P}  \; \; .
\end{split}
\end{equation}
We see that in any case: $ \dot{E}_{B} \;  < \;  - \dot{E}_{R}$, so that there is enough rotational energy to power the
pulsar emission.

\section{\normalsize{RADIO DEATH LINE}}
\label{death}
As discussed in previous Section, until $ \dot{E}_{B} \; < \; - \, \dot{E}_{R}$ the  rotation power loss  sustains the
pulsar emission. We  have already shown that this explain the pulsar activity for pulsars with inferred magnetic fields
well above the critical field $B_{QED}$. In this Section we explain why anomalous $X$-ray pulsars and soft gamma repeaters
are radio quiet pulsars.
\begin{figure}[th]
   \centering
\includegraphics[width=0.9\textwidth,clip]{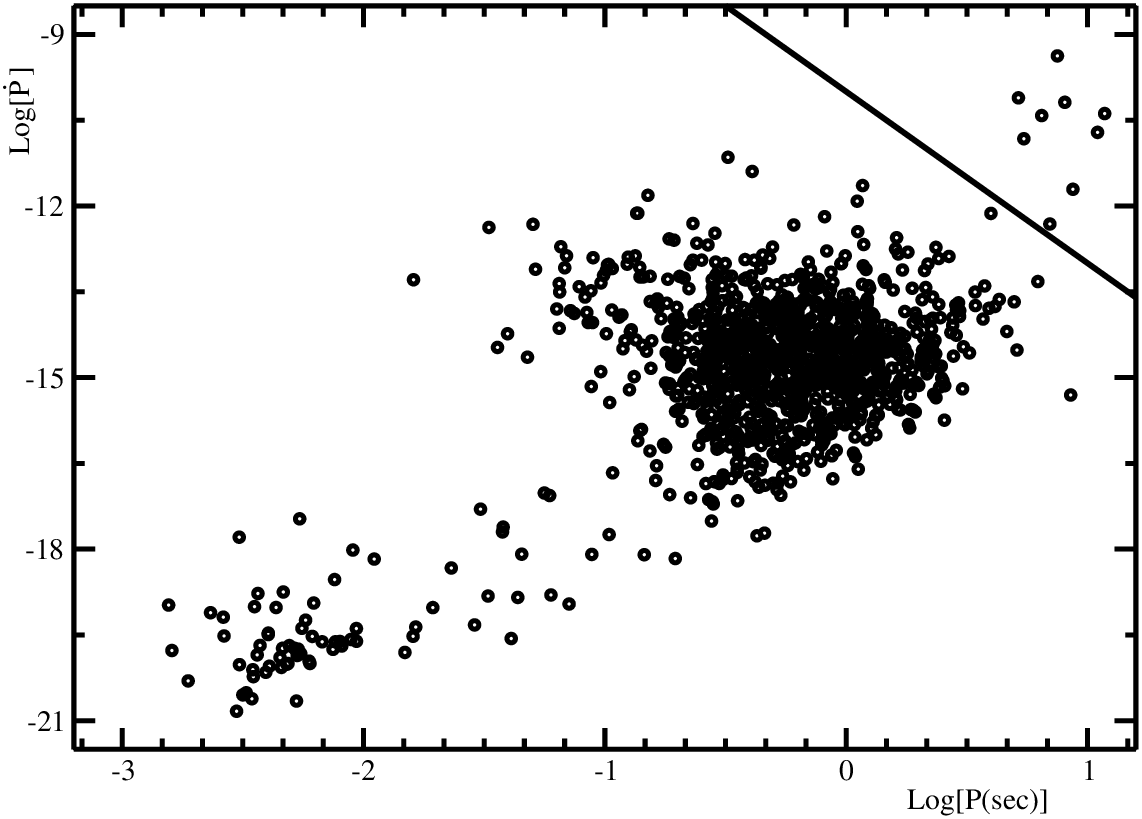}
\caption{\label{fig_2}
Period derivative  plotted versus stellar period for 1194 pulsars taken from the ATNF Pulsar Catalog (\cite{ATNF}). Full
 line corresponds to Eq.~(\ref{death-log})}
\end{figure}
When $ \dot{E}_{B} \; \geq \; - \, \dot{E}_{R}$ all the rotation energy is stored into the increasing magnetic field and
the pulsar emission is turned off. As a consequence pulsars with strong enough magnetic fields are radio quiet.
Accordingly we see that the condition:
\begin{equation}
\label{radio-death}
 \dot{E}_{B} \; = \; - \; \dot{E}_{R} \;  \; .
\end{equation}
is able to distinguish rotation powered pulsars from magnetic powered pulsars. Now, using (\cite{manchester:1977})
\begin{equation}
\label{mag-brake}
 B_S \; \simeq \; \sqrt{\frac{3 \, I \, P \, \dot{P}}{8 \, \pi^2 \, R^6}}  \; \; \; ,
\end{equation}
we recast Eq.~(\ref{radio-death}) into:
\begin{equation}
\label{r-death}
 P^3 \, \dot{P} \; = \; 16 \, \pi^4 \, R^3   \; \; \; .
\end{equation}
Using the canonical radius $R \simeq 10 \, Km$, we get:
\begin{equation}
\label{death-log}
 3 \; \log (P) \, + \, \log (\dot{P})  \; \simeq  \; - \, 10  \; \; \; .
\end{equation}
Equation~(\ref{death-log}) is a straight line,  plotted in Fig.~\ref{fig_2}, in the $\log(P)-\log(\dot{P})$ plane. In
Figure.~\ref{fig_2} we have also displayed 1194 pulsars taken from the ATNF Pulsar Catalog (\cite{ATNF}). We see that
rotation powered pulsars, ranging from millisecond pulsars up to radio pulsars, do indeed lie below our
Eq.~(\ref{death-log}). Note that in Fig.~\ref{fig_2} the recently detected high magnetic field pulsars are not included.
However,  we have already argued in previous Section that these pulsars have spin parameters which indicate that these
pulsars are rotation powered. On the other hand, Fig.~\ref{fig_2} shows that all soft gamma-ray repeaters and anomalous
$X$-ray pulsars in the  ATNF Pulsar Catalog lie above our line Eq.~(\ref{death-log}). In particular, in Fig.~\ref{fig_2}
the pulsar above and nearest to the line Eq.~(\ref{death-log}) corresponds to AXP 1E 2259+586. So that, we see that our
radio dead line, Eq.~(\ref{death-log}), correctly predicts that AXP 1E 2259+586 is not a radio pulsar even though the
magnetic field is lower than that in radio pulsars PSR J1718-3718 and PSR J1847-0130. We may conclude that pulsars above
our dead line are magnetars, i.e. magnetic powered pulsars. The  emission properties of magnetars are quite different from
rotation powered pulsars: the emission from magnetars  consists in thermal blackbody radiation form the surface.
\section{\normalsize{GLITCHES IN MAGNETARS}}
\label{glitches}
\begin{figure}[th]
   \centering
\includegraphics[width=0.9\textwidth,clip]{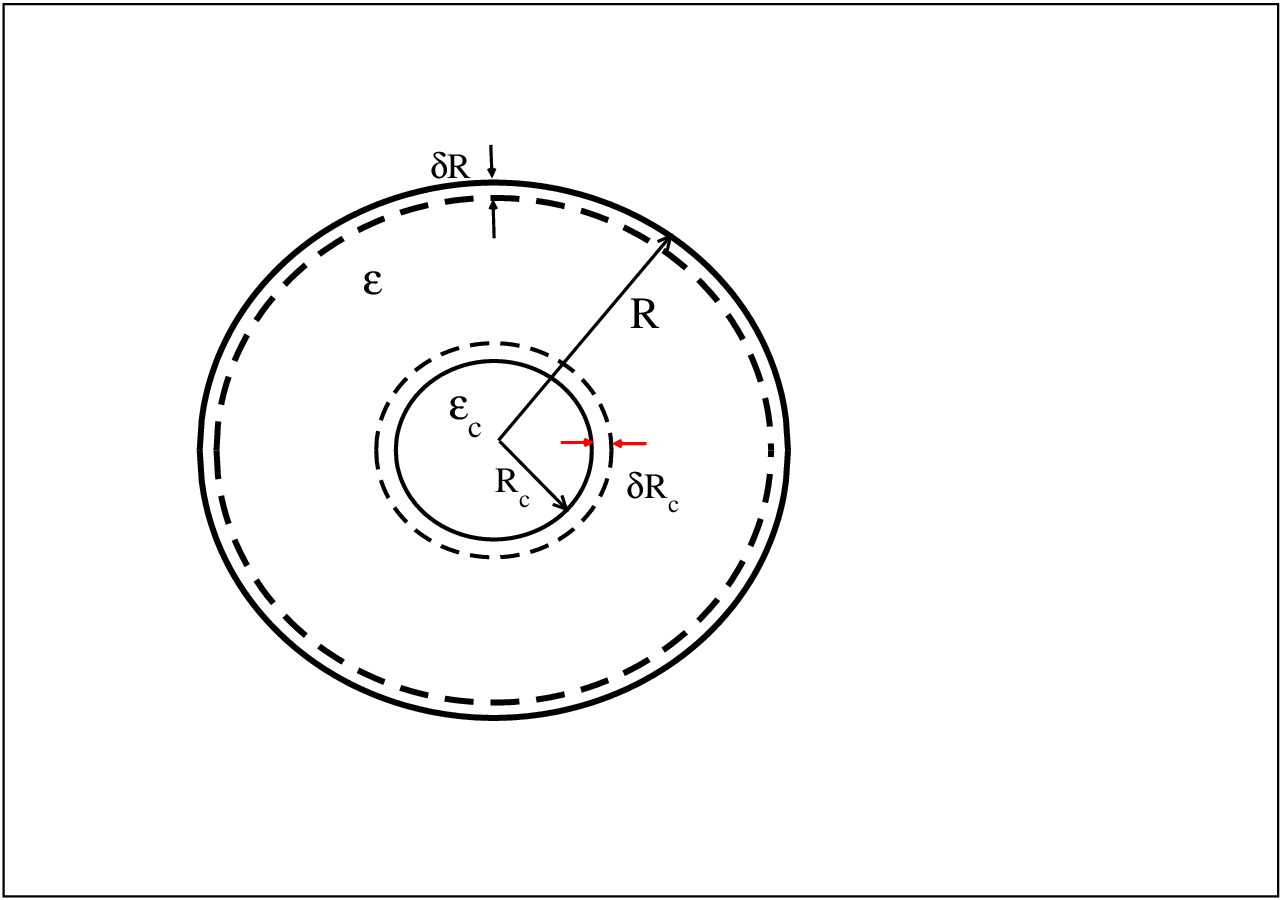}
\caption{\label{fig_3}
Schematic view of the interior of a p-star. $R_c$ and $R$ represent the inner core and stellar radii respectively;
$\varepsilon_c$ is the energy density of the inner core, $\varepsilon$ is the energy density outside the core.}
\end{figure}

The origin of the dipolar magnetic field in p-stars is due to the inner core uniformly magnetized. In Figure~\ref{fig_3}
we display a schematic view of the interior of a p-star.
The presence of the inner core uniformly magnetized leads to well defined glitch mechanism in p-stars. Indeed, dissipative
effects, which are more pronounced in young stars, tend to decrease the strength of the core magnetic field. On the other
hand, when $B_c$ decreases due to dissipation effects, then the magnetic flux locally decreases and, according to Lenz's
law, induces a current which resists to the reduction of the magnetic flux. This means that some quarks must flow into the
core by jumping onto higher Landau levels. In other words, the core radius must increase. Moreover, due to very high
conductivity of  quarks in the core, we have:
\begin{equation}
\label{magn-flux}
B_c \; \; R_c^2  \simeq \; constant \; ,
\end{equation}
which implies:
\begin{equation}
\label{magn-var-1}
\frac{\delta \, B_c}{B_c} \; + \; 2 \; \frac{\delta \, R_c}{R_c} \; \simeq \; 0 \; ,
\end{equation}
or
\begin{equation}
\label{magn-var-2}
\frac{\delta \, R_c}{R_c} \; \simeq \; - \; \frac{1}{2} \; \frac{\delta \, B_c}{B_c}  \; .
\end{equation}
Equation~(\ref{magn-var-2}) confirms that to the decrease of the core magnetic field, $\delta B_c < 0$, it corresponds an
increase of the inner core radius $\delta R_c > 0$. This sudden variation of the radius of the inner core leads to
glitches. Indeed, it is straightforward to show that  the magnetic moment:
\begin{equation}
\label{mag-mom}
 \emph{m} \; = \; B_S \; R^3  \; =  \; \; B_c \; R_c^3  \;
\end{equation}
where we used Eq.~(\ref{magn-core-surf}), must increase in the glitch.  Using Eq.~(\ref{magn-var-2}), we get:
\begin{equation}
\label{mom-var}
\frac{\delta \, \emph{m}}{\emph{m}} \; =  \; \frac{\delta \, B_c}{B_c} \; + \; 3 \; \frac{\delta \, R_c}{R_c} \; \simeq \;
 \frac{\delta \, R_c}{R_c} \; > \;0 \; ,
\end{equation}
Another interesting consequence of the glitch is that the stellar radius $R$ must decrease, i.e. the star contracts. This
is an inevitable consequence of the increase of the inner core, which is characterized by an energy  density
$\varepsilon_c$ higher then the outer core density  $\varepsilon$. As a consequence the  variation of radius is negative:
$\delta R < 0$ (see Fig.~\ref{fig_3} ). In radio pulsar, where the magnetic energy can be neglected, conservation of the
mass leads to:
\begin{equation}
\label{rad-var-1}
\frac{\delta \, R}{R} \; \simeq \; - \; \frac{\varepsilon_c - \varepsilon}{\overline{\varepsilon}}  \; \left  ( \frac{R_c
}{ R} \right )^3
 \;   \frac{\delta \, R_c}{R_c} \; ,
\end{equation}
where $\overline{\varepsilon}$ is the average energy density. In general, we may assume that $\frac{\varepsilon_c -
\varepsilon}{\overline{\varepsilon}}$ is a constant of order unity. So that Eq.~(\ref{rad-var-1}) becomes:
\begin{equation}
\label{rad-var-2}
\frac{\delta \, R}{R} \; \simeq \; - \;  \left ( \frac{R_c }{ R} \right )^3 \;  \frac{\delta \, R_c}{R_c} \; .
\end{equation}
Note that the ratio $ \left  ( \frac{R_c }{ R} \right )^3$ can be estimate from Eq.~(\ref{magn-core-surf}) once the
surface magnetic field is known. We find that, even for magnetars,  $\left  ( \frac{R_c }{ R} \right )^3$ is of order
$10^{-2}$ or less. So that our Eqs.~(\ref{magn-core-surf}) and~(\ref{rad-var-2}) show that:
\begin{equation}
\label{var-rad-cor}
 \delta R \; < \;  0 \; \; \; ,  \; \; \; - \; \frac{\delta \, R}{R} \; \ll \;
 \frac{\delta \, R_c}{R_c} \; \; .
\end{equation}
As is well known, because the external magnetic braking torque, pulsars slow down according to (e.g. see Manchester \&
Taylor 1977):
\begin{equation}
\label{brak-eq}
 - \; \dot{\nu} \; \propto \;   \emph{m}^2 \; I^{-1} \; \nu^3 \; .
\end{equation}
So that:
\begin{equation}
\label{delta-brak-eq}
 \frac{\delta \dot{\nu} }{\dot{\nu}} \; = \; 2 \; \frac{\delta \emph{m}}{ \emph{m}}  \; - \; \frac{\delta I}{I} \;
 + \; 3 \;  \frac{\delta \nu}{\nu} \; \;  .
\end{equation}
From conservation of angular momentum we have:
\begin{equation}
\label{i-nu}
| \frac{\delta I}{I} | \; \; \simeq \; \;
 | \frac{\delta \nu}{\nu} | \; \;  .
\end{equation}
Moreover, from observational data it turns out that:
\begin{equation}
\label{nudot-nu}
| \frac{\delta  \dot{\nu}}{ \dot{\nu}} | \; \; \gg \; \;
 | \frac{\delta \nu}{\nu} | \; \;  ,
\end{equation}
so that Eq.~(\ref{delta-brak-eq}) becomes:
\begin{equation}
\label{delta-brak-eq-2}
 \frac{\delta \dot{\nu} }{\dot{\nu}} \; \simeq \; 2 \; \frac{\delta \emph{m}}{ \emph{m}}  \; \simeq \;
 2 \; \frac{\delta \, R_c}{R_c} \; \simeq \; - \; \frac{\delta \, B_c}{B_c} \; \; ,
\end{equation}
where we used Eqs.~(\ref{mom-var}) and~(\ref{magn-var-2}). Equation~(\ref{delta-brak-eq-2}) does  show that the variation
of the radius of the inner core leads to a glitch. \\
In rotation powered pulsar, starting from Eq.~(\ref{rad-var-1}) one can show that $ | \frac{\delta \nu}{\nu} | \simeq |
\frac{\delta \, R}{R} |$. So that, taking into account Eq.~(\ref{var-rad-cor}) we recover the phenomenological relation
Eq.~(\ref{nudot-nu}). A full account of glitches in radio pulsar will be presented elsewhere. Glitches in magnetars are
considered in the next Section, where we
show that, indeed,  Eqs.~(\ref{var-rad-cor}) and~(\ref{nudot-nu}) hold even in magnetars. \\
The most dramatic effect induced by glitches in magnetars is the release of a huge amount of magnetic energy in the
interior of the star and into the magnetosphere. To see this, let us consider the energy stored into the magnetic field in
the interior of the magnetar. We have:
\begin{equation}
\label{mag-ener-int}
E_{B}^{\emph{int}} \; = \; \frac{1}{6} \; R_c^3 \; B_c^2 \; + \; \frac{1}{8 \, \pi} \;  \int_{R_c}^{R} \;  \; d^3r \;
\left [ B_c \; \left (\frac{R_c}{r} \right )^3 \right ]^2 \;  \; \; ,
\end{equation}
where the first term is the energy stored into the core where the magnetic field is uniform. The variation of the magnetic
energy Eq.~(\ref{mag-ener-int}) caused by a glitch is easily evaluated. Taking into account Eq.~(\ref{magn-var-2}) and
$\left  ( \frac{R_c }{ R} \right )^3 \ll 1$, we get:
\begin{equation}
\label{var-mag-ener-int}
\delta \, E_{B}^{\emph{int}} \; \simeq \; - \; \frac{1}{3} \; R^3 \; B_S^2 \; \left  ( \frac{R}{ R_c} \right )^3 \;
\frac{\delta  R_c}{R_c} \; - \; \frac{1}{2} \;   R^3 \; B_S^2 \;  \frac{\delta  R_c}{R_c} \; \simeq \; - \; \frac{1}{3} \;
R^3 \; B_S^2 \; \left  ( \frac{R}{ R_c} \right )^3 \; \frac{\delta  R_c}{R_c}  \; .
\end{equation}
Equation~(\ref{var-mag-ener-int}) shows that there is a decrease of the magnetic energy. So that after a glitch in
magnetars a huge magnetic energy is released  in the interior of the star. We shall see that this energy is enough to
sustain the quiescent emission. On the other hand, the glitch induces also a sudden variation of the magnetic energy
stored into the magnetosphere. Indeed, from Eq.~(\ref{magn-core-surf}) we find:
\begin{equation}
\label{magn-surf-var}
\frac{\delta B_S}{B_S} \; = \; \frac{\delta B_c}{B_c} \; + \; 3 \; \frac{\delta R_c}{R_c} \;  -\; 3 \; \frac{\delta R}{R}
\; \simeq \; \frac{\delta R_c}{R_c} \;  -\; 3 \; \frac{\delta R}{R} \; \simeq \; \frac{\delta R_c}{R_c} \; > \; 0 \; \; .
\end{equation}
Thus, the magnetic energy stored into the magnetosphere:
\begin{equation}
\label{mag-ener-ext}
 E_{B}^{\emph{ext}} \; = \; \frac{1}{8 \, \pi} \;  \int_{R}^{\infty} \;  \; d^3r \;
\left [ B_S \; \left (\frac{R}{r} \right )^3 \right ]^2 \; = \; \frac{1}{6 } \; B_S^2 \; \; R^3 \; \; ,
\end{equation}
increases by:
\begin{equation}
\label{var-mag-ener-ext}
\delta  E_{B}^{\emph{ext}} \; = \;  \frac{1}{3} \; R^3 \; B_S^2 \; \left  ( \frac{\delta B_S}{ B_S} \; + \;
 \frac{3}{2}  \frac{\delta R}{R} \right ) \; \simeq \; \frac{1}{3} \; R^3 \; B_S^2 \; \frac{\delta B_S}{ B_S} \;
  > \;  0  \; \; .
\end{equation}
This magnetic energy is directly injected  into the magnetosphere, where it is dissipated by well defined physical
mechanism discussed in Section~\ref{bursts}, and it is responsible for bursts in soft gamma-ray repeaters and anomalous
$X$-ray pulsars. \\
To summarize,  in this Section we have found that dissipative phenomena in the inner core of a p-star tend to decrease the
strength of the core magnetic field. This, in turn, results in an increase of the radius of the core $\delta R_c > 0$, and
in a contraction of the surface of the star,  $\delta R < 0$. We have also shown that the glitch releases an amount of
magnetic energy in the interior of the star and injects magnetic energy  into the magnetosphere, where it is completely
dissipated. Below we will argue that these magnetic glitches are responsible for the quiescent emission and bursts in
gamma-ray repeaters and anomalous $X$-ray pulsars. Interestingly enough, in Cheng et al. (1995)  it was shown that SGR
events and earthquakes share several distinctive statistical properties, namely: power-law energy distributions,
log-symmetric waiting time distributions, strong correlations between waiting times of successive events, and weak
correlations between waiting times and intensities. These statistical properties of bursts can be easily understood if
bursts originate by  the release of a small amount of energy from a reservoir of stored energy. As a matter of fact, in
our theory the burst activity is accounted for by the release of a tiny fraction of magnetic energy stored in magnetars.
Even for giant bursts we find that the released energy is a few per cent of the magnetic energy. Moreover, Hurley et al.
(1994) noticed that there is a significantly  statistical similarity between the bursts from  SGR 1806-20 and
microglitches  observed from the Vela pulsar with $ |\frac{\delta \nu}{\nu}| \thicksim 10^{-9}$. So that we see that these
early statistical studies of bursts are in agreement with our theory for bursts in magnetars. Even more, we shall show
that after a giant glitch there is an intense burst activity quite similar to the settling earthquakes following a strong
earthquake.
\subsection{\normalsize{BRAKING GLITCHES}}
\label{braking}
In Section~\ref{glitches} we found that magnetic glitches in p-stars lead to:
\begin{equation}
\label{delta-dot-nu}
 \frac{\delta \dot{\nu} }{\dot{\nu}} \; \; \simeq \; \; - \; \frac{\delta  B_c}{B_c} \; > \; 0 \; \; .
\end{equation}
Since there is variation of both the  inner core and the stellar radius, the moment of inertia of the star undergoes a
variation of $\delta I$. It is easy to see that the increase of the inner core leads to an increase of the moment of
inertia $I$; on the other hand, the reduction of the stellar radius implies $\delta I < 0$. In radio pulsar, where, by
neglecting the variation of the magnetic energy, the conservation of the mass leads to Eq.~(\ref{rad-var-2}), one can show
that:
\begin{equation}
\label{var-I-radio}
 \frac{\delta I}{I}  \; \; \simeq \; \; \frac{\delta  R}{R} \; < \; 0 \; \; .
\end{equation}
Moreover, from conservation of angular momentum:
\begin{equation}
\label{cons-ang}
 \frac{\delta I}{I}  \; \; = \; - \; \frac{\delta \nu}{\nu} \; \; ,
\end{equation}
it follows:
\begin{equation}
\label{radio-glitch}
0 \; < \;  \; \frac{\delta \nu}{\nu} \; \;  \simeq \; \; - \; \frac{\delta  R}{R} \; \ll \; - \; \frac{\delta  B_c}{B_c}
\; \;  \simeq \; \; \frac{\delta \dot{\nu} }{\dot{\nu}} \; \; .
\end{equation}
For magnetars, namely p-stars with super strong magnetic field, the variation of magnetic energy cannot be longer
neglected. In this case, since the magnetic energy decreases, we have that the surface contraction in magnetars is smaller
than in radio pulsars. That means that Eq.~(\ref{var-rad-cor}) holds even for magnetars. Moreover, since in radio pulsars
we known that:
\begin{equation}
\label{radio-glitch-max}
 \frac{\delta \nu}{\nu} \;  \;  = \; - \;  \frac{\delta I}{I}  \;\simeq \; \; - \; \frac{\delta  R}{R} \; \lesssim \;
 10^{-6} \; \; ,
\end{equation}
we see that in magnetars the following bound must hold:
\begin{equation}
\label{R-bound}
 - \; \frac{\delta  R}{R} \; \lesssim \;
 10^{-6} \; \; .
\end{equation}
As a consequence we may write:
\begin{equation}
\label{var-I-magnetar}
 \frac{\delta I}{I}  \; \; = \;  \; \left( \frac{\delta I}{I} \right )_{core} \; \; +  \; \;
 \left( \frac{\delta I}{I} \right )_{surf} \; \; , \; \;
 \left( \frac{\delta I}{I} \right )_{surf} \simeq \; \; \frac{\delta  R}{R} \; < \; 0 \; .
\end{equation}
As we show in a moment, the variation of the moment of inertia  induced by the core is positive. So that if the core
contribution overcomes the surface contribution to $\delta I$ we have a \emph{braking glitch} where $- \frac{\delta
\nu}{\nu} = \frac{\delta P}{P} > 0$. \\
We believe that the most compelling evidence in support to our proposal comes from the anomalous $X$-ray pulsar AXP 1E
2259+586. As reported in Woods et al. (2004), the timing data showed that a large glitch occurred  in  AXP 1E 2259+586
coincident with the 2002 June giant burst. Remarkably, at the time of the giant flare on 1998 August 27, the soft gamma
ray repeater  SGR 1900+14 displayed a discontinuous spin-down consistent with a braking glitch (\cite{Woods:1999}). Our
theory is able to explain why AXP 1E 2259+586 displayed a normal glitch, while  SGR 1900+14 suffered a braking glitch. To
see this, we recall the spin-down parameters of these pulsars:
\begin{equation}
\label{spin-par-magnetar}
\begin{split}
  SGR \; 1900+14   &   \;  \; \; P  \;  \simeq \; 5.16 \; sec \;  ,  \;  \dot{P}  \;  \simeq \; 1.1 \; 10^{-10}
                       \; ,  \; B_S \;  \simeq \;  7.4 \; 10^{14} \; Gauss \; \; ,  \\
  AXP \; 1E 2259+586  &  \;  \; \;  P \;  \simeq \; 6.98 \; sec \;  ,  \;  \dot{P}  \;  \simeq \; 2.0 \; 10^{-14}
                         \; , \; B_S \;  \simeq \;  1.2 \; 10^{13} \; Gauss \; \; .
\end{split}
\end{equation}
For canonical magnetars with  $M \, \simeq 1.4 \, M_{\bigodot}$ and radius  $R \, \simeq 10  \, Km$, we have $\sqrt{gH}
\simeq  0.55 \; GeV$. So that, using $ 1 \, GeV^2 \, \simeq \, 5.12 \, 10^{19} \, Gauss$, we rewrite
Eq.~(\ref{magn-core-surf}) as:
\begin{equation}
\label{magn-surf-num}
B_S\; \; \simeq \; \; 1.54 \;  10^{16}  \; \left (\frac{R_c}{R } \right )^3 \;  \; Gauss \; \; .
\end{equation}
Combining Eqs.~(\ref{spin-par-magnetar}) and~(\ref{magn-surf-num}) we get:
\begin{equation}
\label{R_c-R}
\begin{split}
 SGR \; 1900+14   &   \; \; \;\; \; \;  \;  \; \left (\frac{R_c}{R } \right )^3 \;  \; \simeq  \; 4.81 \; 10^{-2}
                          \; \; , \\
 AXP \; 1E 2259+586 &  \; \; \;\;\; \; \;  \;   \left (\frac{R_c}{R } \right )^3 \; \; \simeq   \; 0.78 \; 10^{-3}
                          \; \; .
\end{split}
\end{equation}
According to Eqs.~(\ref{delta-dot-nu}), ~(\ref{cons-ang}) and~(\ref{var-I-magnetar}), to evaluate the sudden variation of
the frequency and frequency derivative,  we need $\delta  R$ and $\delta  R_c$. These parameters can be estimate from the
energy released during the giant bursts. In the case of AXP 1E 2259+586,  the giant  2002 June burst followed an intense
burst activity which lasted for about one year. Woods et al. (2004), assuming a distance of $3 \; kpc$ to 1E 2259+586,
measured an energy release of $2.7 \, 10^{39} \, ergs$ and  $2.1 \,  10^{41} \, ergs$ for the fast and slow decay
intervals, respectively. Due to this uncertainty, we conservatively estimate the energy released during the giant burst to
be:
\begin{equation}
\label{GB-2259}
  AXP \; 1E 2259+586   \; \; \; \;  \;  E_{\emph{burst}}  \;  \;    \; \; \simeq   \; 1.0 \; 10^{40}
                                             \; \; ergs \; \; .
\end{equation}
On 1998 August 27, a giant burst from the soft gamma ray repeater  SGR 1900+14 was recorded. The estimate energy released
was:
\begin{equation}
\label{GB-1900}
  SGR \; 1900+14   \; \; \; \;  \;  E_{\emph{burst}}  \;  \;    \; \; \simeq   \; 1.0 \; 10^{44}
                                             \; \; ergs \; \; .
\end{equation}
As we have already discussed in Sect.~\ref{glitches}, the energy released during a burst in magnetars is given by the
magnetic energy directly injected and dissipated into the magnetosphere, Eq.~(\ref{var-mag-ener-ext}). We rewrite
Eq.~(\ref{var-mag-ener-ext}) as
\begin{equation}
\label{var-mag-ener-ext-num}
\delta  E_{B}^{\emph{ext}}  \; \simeq \; \frac{1}{3} \; R^3 \; B_S^2 \; \frac{\delta B_S}{ B_S} \;
                             \simeq \;  2.6 \; 10^{44} \; ergs  \; \left ( \frac{B_S}{10^{14} \, Gauss} \right )^2
                             \; \frac{\delta B_S}{ B_S}  \; \; .
\end{equation}
So that, combining Eqs.~(\ref{var-mag-ener-ext-num}), (\ref{GB-1900}), (\ref{GB-2259})  and (\ref{spin-par-magnetar}) we
get:
\begin{equation}
\label{burst-param-1900-2259}
\begin{split}
 SGR \; 1900+14   &   \; \; \;\; \; \;  \;  \;  \frac{\delta B_S}{ B_S}  \; \simeq \; \frac{\delta R_c}{R_c} \;
                           \simeq \; 0.70 \; 10^{-2}    \; \;  \; , \\
AXP \; 1E 2259+586 &    \; \; \;\; \; \;  \;  \;  \frac{\delta B_S}{ B_S}  \; \simeq \; \frac{\delta R_c}{R_c} \;
                            \simeq \; 0.27 \; 10^{-2}    \; \; \; .
\end{split}
\end{equation}
Thus, according to Eq.~(\ref{delta-dot-nu}) we may estimate the sudden variation of $\dot{\nu}$:
\begin{equation}
\label{GB-delta-dot-nu}
 \frac{\delta \dot{\nu} }{\dot{\nu}} \; \; \simeq \; \;  2 \; \frac{\delta \, R_c}{R_c} \; \; \thicksim \; \; 10^{-2}
  \; \;,
\end{equation}
for both glitches. On the other hand we have:
\begin{equation}
\label{B-delta-I_c}
 \left( \frac{\delta I}{I} \right )_{core} \; \;  \simeq \; \; \frac{15}{2} \;
  \frac{\varepsilon_c - \varepsilon}{\overline{\varepsilon}}
  \; \left  ( \frac{R_c}{ R} \right )^5 \;   \frac{\delta \, R_c}{R_c} \; \; \simeq \; \;
  \frac{15}{2} \; \; \left  ( \frac{R_c}{ R} \right )^5 \;   \frac{\delta \, R_c}{R_c} \; \; ,
\end{equation}
leading to:
\begin{equation}
\label{B-delta-I_c-1900-2259}
\begin{split}
SGR \; 1900+14   &   \; \; \;\; \; \;  \;  \;   \left( \frac{\delta I}{I} \right )_{core}
                           \simeq \;  3.34 \; \; 10^{-4}    \; \;  \; , \\
AXP \; 1E 2259+586 &    \; \; \;\; \; \;  \;  \;  \left( \frac{\delta I}{I} \right )_{core}
                            \simeq \;  1.34 \; \; 10^{-7}      \; \; \; .
\end{split}
\end{equation}
On the other hand, we expect that during the giant glitch $\left( \frac{\delta I}{I} \right )_{surf} \thicksim 10^{-6}$.
As a consequence, for AXP  1E 2259+586 the core contribution is negligible with respect to  the surface contribution to
$\delta I$. In other words, AXP  1E 2259+586 displays a normal glitch with $\frac{\delta \nu}{\nu} \thicksim 10^{-6}$. On
the contrary, Eq.~(\ref{B-delta-I_c-1900-2259}) indicates that SGR  1900+14 suffered a braking glitch with $\frac{\delta
\nu}{\nu} \thicksim -  3.34 \; 10^{-4}$ giving:
\begin{equation}
\label{B-braking-1900}
 \Delta \; P  \; \;  \simeq \; \; 1.72  \; \; 10^{-3} \; \; sec \; \; .
\end{equation}
We see that our theory is in agreement with observations, for  a glitch of size $\frac{\delta \nu}{\nu} = 4.24(11) \;
10^{-6}$ was observed in AXP  1E 2259+586 which preceded the burst activity (\cite{Woods:2004a}). Moreover, our theory
predicts a sudden increase of the spin-down torque  according to Eq.~(\ref{GB-delta-dot-nu}). In Woods et al. (2004) it is
pointed out that it was not possible to give a reliable estimate of the variation of the frequency derivative since the
pulse profile was undergoing large changes, thus compromising the phase alignment with the pulse profile template. Indeed,
as discussed in Sect.~\ref{2259}, soon after the giant burst AXP 1E 2259+586 suffered an intense burst activity. Now,
according to our theory, during the burst activity there is both a continuous injection of magnetic energy into the
magnetosphere and variation of the magnetic torque explaining the anomalous timing noise observed in  1E 2259+586. In
addition, Woods et al. (1999b) reported a gradual increase of the nominal spin-down rate and a discontinuous spin down
event associated with the 1998 August 27 flare from SGR 1900+14. Extrapolating the long-term trends found before and after
August 27, they found evidence of a braking glitch with $\Delta P \,  \simeq \,  0.57(2)  \; 10^{-3} \;  sec$. In view of
our theoretical uncertainties, the agreement with our Eq.~(\ref{B-braking-1900}) is rather good.
To our knowledge the only attempt to explain within the standard model the braking glitch observed in SGR  1900+14 is done
in Thompson et al. (2000) where it is suggested that violent August 27 event involved a glitch. The magnitude of the
glitch was estimated by scaling to the largest glitches in young, active pulsars with similar spin-down ages and internal
temperature. In this way they deduced the estimate $|\frac{\Delta P}{P}| \thicksim 10^{-5}$ to $10^{-4}$. However, this
explanation  overlooks the well known fact that radio pulsars display normal glitches and no braking glitches. Moreover,
there is no arguments to explain why AXP 1E 2259+586 displayed a normal glitch instead of a braking glitch. \\
Let us conclude this Section by briefly discussing the 2004 December 27 giant flare from SGR  1806-20. During this
tremendous outburst SGR  1806-20 released a huge amount of energy, $ E_{\emph{burst}} \thicksim 10^{46} \; ergs$. Using
the spin-down parameters reported in Mereghetti et al. (2005a):
\begin{equation}
\label{spin-par-1806}
 SGR \; 1806-20    \;  \; \; P  \;  \simeq \; 7.55 \; sec \;  ,  \;  \dot{P}  \;  \simeq \; 5.5 \; 10^{-10}
                       \; ,  \; B_S \;  \simeq \;  2.0 \; 10^{15} \; Gauss \; \; ,
\end{equation}
we find:
\begin{equation}
\label{burst-param-1806}
 SGR \; 1806-20  \; \;\; \; \;   \frac{\delta B_S}{ B_S}  \; \simeq \; \frac{\delta R_c}{R_c} \;
                           \simeq \; 9.6 \; 10^{-2}    \; \; .
\end{equation}
Thus, we predict that SGR  1806-20 should display a gigantic braking glitch with $\frac{\Delta P}{P} \simeq
 2.4 \; 10^{-2}$, or :
\begin{equation}
\label{B-braking-1806}
 \Delta \; P  \; \;  \simeq \; \; 1.8  \; \; 10^{-1} \; \; sec \; \; .
\end{equation}
\subsection{\normalsize{QUIESCENT LUMINOSITY}}
\label{quiescent}
The basic mechanism to explain the quiescent $X$-ray emission in our magnetars is the internal dissipation of magnetic
energy. Our mechanism is basically the same as in the standard magnetar model based on neutron star (\cite{Duncan:1996}).
Below we shall compare our proposal with the standard theory. In Section~\ref{glitches} we showed that during a glitch
there is a huge amount of magnetic energy released into the magnetar:
\begin{equation}
\label{delta-ener-int}
 - \; \delta \, E_{B}^{\emph{int}}  \; \; \simeq \;  \; \frac{1}{3} \;
R^3 \; B_S^2 \; \left  ( \frac{R}{ R_c} \right )^3 \; \frac{\delta  R_c}{R_c}  \; .
\end{equation}
As in previous Section, we use SGR 1900+14 and  AXP 1E 2259+586 as prototypes for soft gamma ray repeaters and anomalous
$X$-ray pulsars, respectively. Using the results of Sect.~\ref{braking}, we find:
\begin{equation}
\label{delta-ener-1900-2259}
\begin{split}
SGR \; 1900+14   &   \; \; \;\; \; \;  \;  \; - \; \delta \, E_{B}^{\emph{int}}  \; \; \simeq \; \;  3.0 \; \; 10^{47}
                          \;  \;  erg  \; \;  \frac{\delta  R_c}{R_c} \;  \; , \\
AXP \; 1E 2259+586 &   \; \; \;\; \; \;  \;  \; - \; \delta \, E_{B}^{\emph{int}}  \; \; \simeq \; \;  4.8 \; \;
                           10^{45}  \;  \;  erg  \; \;  \frac{\delta  R_c}{R_c} \;  \; .
\end{split}
\end{equation}
This release of magnetic energy is dissipated leading to observable surface luminosity. To see this, we need a thermal
evolution model which calculates the interior temperature distribution. In the case of neutron stars such a calculation
has been performed in VanRiper (1991), where it is showed that the isothermal approximation is a rather good approximation
in the range of inner temperatures of interest. The equation which determines the thermal history of a p-star has been
discussed in Cea (2004a) in the isothermal approximation:
\begin{equation}
\label{cooling}
 C_V \; \frac{d T}{d t}
\; = \; - \; (L_{\nu} \; + \; L_{\gamma}) \; ,
\end{equation}
where $L_{\nu}$ is the neutrino luminosity, $L_{\gamma}$ is the photon luminosity and $C_V$ is the specific heat. Assuming
blackbody photon emission from the surface at an effective surface temperature $T_S$ we have:
\begin{equation}
\label{blackbody}
 L_{\gamma} \; = \; 4 \, \pi \, R^2 \, \sigma_{SB} \, T_S^4 \; ,
\end{equation}
where $\sigma_{SB}$ is the $Stefan-Boltzmann$ constant. In  Cea (2004a)  we assumed that the surface and interior
temperature were related by:
\begin{equation}
\label{surface}
 \frac{T_S}{T} \; = \; 10^{-2} \; a  \; \; .
\end{equation}
Equation~(\ref{surface}) is relevant for a p-star which is not bare, namely for p-stars which are endowed with a thin
crust. The vacuum gap between the core and the crust, which is of order of several hundred \emph{fermis}, leads to a
strong suppression of the surface temperature with respect to the core temperature. The precise relation between $T_S$ and
$T$ could be obtained by a careful study of the crust and core thermal interaction. In any case, our phenomenological
relation Eq.~(\ref{surface}) allows a wide variation of $T_S$, which encompasses the neutron star relation (see, for
instance, \cite{gudmundsson:1983}). Moreover, our cooling curves display a rather weak dependence on the parameter $a$ in
Eq.~(\ref{surface}). Since we are interested in the quiescent luminosity, we do not need to known the precise value of
this parameter.  So that, in the following we shall assume $a \thicksim 1$. In other words, we assume:
\begin{equation}
\label{surface-new}
  T_S \;  \;  \simeq \; \; 10^{-2} \; \; T \; \; \; .
\end{equation}
Obviously, the parameter $a$ is relevant to evaluate the surface temperature once the core temperature is given. Note
that, in the relevant range of core temperature $ T \thicksim 10^{8}$ $\, {}^\circ K$, our Eq.~(\ref{surface-new}) is
practically identical to the parametrization adopted in Duncan \& Thompson (1996) within the standard magnetar model:
\begin{equation}
\label{Temp-D-T}
  T_S \;  \;  \simeq \; \; 1.3 \; \; 10^6 \; {}^\circ K \; \; \left ( \frac{T}{10^8 \, {}^\circ K} \right )^{\frac{5}{9}}
                \; \; \; .
\end{equation}
The neutrino luminosity $L_{\nu}$ in Eq.~(\ref{cooling}) is given by the direct $\beta$-decay quark reactions, the
dominant cooling processes by neutrino emission. From the results in Cea (2004a), we find:
\begin{equation}
\label{luminosity}
 L_{\nu} \; \; \simeq \; \; 1.22 \; \; 10^{36} \; \;  \frac{erg}{s} \; \;  T_9^8 \; \; \;  ,
\end{equation}
where $T_9$ is the temperature in units of $10^9$ $\, {}^\circ K$.  Note that the neutrino luminosity $L_{\nu}$ has the
same temperature dependence as the neutrino luminosity by  modified URCA reactions in neutron stars, but it is more than
two order of magnitude smaller.
From the cooling curves reported in Cea (2004a) we infer that the surface and interior temperature are almost constant up
to time $\tau \thicksim 10^{5} \, years$. Observing that magnetars candidates are rather young pulsar with $\tau_{age}
\lesssim 10^{5} \, years$, we may estimate the average surface luminosities as:
\begin{equation}
\label{Luminosity}
 L_{\gamma} \; \; \simeq \; \frac{ - \; \delta \, E_{B}^{\emph{int}} }{\tau_{age}}  \; \; \; \; \; .
\end{equation}
We assume $ \tau_{age} \thickapprox \tau_c$ for {\it {SGR 1900+14}}. On the other hand, it is known that for AXP 1E
2259+586 $ \; \; \tau_{age} \thicksim 10^3 \, years \ll \tau_c$. So that, we get:

\begin{equation}
\label{lum-1900-2259}
\begin{split}
SGR \; 1900+14   &   \; \; \;\; \; \;  \;  \;  L_{\gamma}  \; \; \simeq \; \; 1.3 \; \; 10^{37}
                          \;  \; \frac{erg}{s}  \; \;  \frac{\delta  R_c}{R_c} \;  \; , \\
AXP \; 1E 2259+586 &   \; \; \;\; \; \;  \; \;  L_{\gamma} \; \; \simeq \; \;  1.5 \; \;
                           10^{35}  \;  \; \frac{erg}{s} \; \;  \frac{\delta  R_c}{R_c} \;  \; .
\end{split}
\end{equation}
We see that it is enough to assume that SGR 1900+14 suffered in the past a glitch with $\frac{\delta  R_c}{R_c} \thicksim
10^{-2}$ to sustain the observed luminosity $ L_{\gamma} \thicksim 10^{35} \; \frac{erg}{s}$ (assuming a distance of about
$10 \; kpc$). In the case of AXP 1E 2259+586, assuming a distance of  about $3 \; kpc$, the observed luminosity is $
L_{\gamma} \thicksim 10^{34} \; \frac{erg}{s}$, so that we infer that this pulsar had suffered in the past a giant glitch
with $\frac{\delta  R_c}{R_c} \thicksim 10^{-1}$, quite similar to the recent SGR 1806-20 glitch.
\\
Let us discuss the range of validity of our approximation. Equation~(\ref{Luminosity}) is valid as long as  $L_{\gamma}$
dominates over  $L_{\nu}$,  otherwise the star is efficiently cooled by neutrino emission and the surface luminosity
saturates to $L_{\gamma}^{max}$. We may quite easily evaluate this limiting luminosity from $L_{\gamma}^{max} \simeq
L_{\nu}$. Using Eq.~(\ref{surface-new}) and $R \simeq 10 \; Km$, we get:
\begin{equation}
\label{Lum-max}
 L_{\gamma}^{max} \; \; \simeq \;  \;   4.2 \; \; 10^{37}  \; \frac{erg}{s}   \; \; \; \; .
\end{equation}
Note that, since our neutrino luminosity is reduced by more than two order of magnitude with respect to neutron stars,
$L_{\gamma}^{max}$ is about two order of magnitude greater than the maximum allowed surface luminosity in neutron stars
(\cite{VanRiper:1991}). Thus, our theory allows to account easily for observed luminosities  up to $10^{36} \;
\frac{erg}{s}$. \\
Let us, finally, comment on the quiescent thermal spectrum in our theory. As already discussed, the origin of the
quiescent emission is the huge release of magnetic energy in the interior of the magnetar. Our previous estimate of the
quiescent luminosities assumed that the interior temperature distribution was uniform. However, due to the huge magnetic
field, the thermal conductivity is enhanced along the magnetic field. This comes out to be the case for both electron and
quarks, since we argued that the magnetic and  chromomagnetic fields are aligned . As a consequence, we expect that the
quiescent spectrum should be parameterized  as two blackbodies with parameter $R_1 \, , \, T_1$ and $R_2 \, , \, T_2$,
respectively. Since the blackbody luminosities $L_{\gamma1}$ and $L_{\gamma2}$ are naturally of the same order, our
previous estimates for the quiescent luminosities are unaffected. Moreover, since the thermal conductivity is enhanced
along the magnetic field, the high temperature blackbody, with temperature $T_2$, originates  from the heated polar
magnetic cups. Thus we have:
\begin{equation}
\label{black-bodies}
\begin{split}
 &  R_1  \; \; \lesssim \;  \; R  \; \; \; , \; \; \; R_2 \lesssim \; 1 \; Km \; \; , \; \; \\
 &  T_2  \; \; >  \; \; T_1 \; \;  \; , \; \; R_1  \; \; >  \; \; R_2 \;  \; , \; \; L_{\gamma1} \; \; \simeq \; \;
    L_{\gamma2} \; \; .
\end{split}
\end{equation}
Note that there is a natural  anticorrelation between  blackbody radii and temperatures.
It is customary to fit  the quiescent spectrum of anomalous $X$-ray pulsars and soft gamma ray repeaters  in terms of
blackbody plus power law. In particular, it is assumed that the power law component extends to energy greater than an
arbitrary cutoff energy $E_{cutoff} \simeq 2 \, KeV$. It is worthwhile to stress that these parameterizations of the
quiescent spectra are in essence phenomenological fits. Indeed, within the standard magnetar model (\cite{Duncan:1996})
the power law should be related to hydromagnetic wind accelerated by Alfven waves. The luminosity of the wind emission
should increase with magnetic field strength as $L_{wind} \simeq L_{PL} \thicksim B_S^2$. On the other hand, the blackbody
luminosities should scale as $B_S^{4.4}$ (\cite{Duncan:1996}). So that the ratio $L_{PL}/L_{BB}$ decreases with increasing
magnetic field strengths, contrary to observations (\cite{Marsden:2001}). In our theory well defined physical arguments
lead to the two blackbody representation of the quiescent spectra, whose parameters are constrained by our
Eq.~(\ref{black-bodies}). As a matter of fact, we have checked in literature that the quiescent spectra of both anomalous
$X$-ray pulsars and soft gamma ray repeaters could be accounted for by two balckbodies. For instance, in Morii et al.
(2003) the quiescent spectrum of AXP 1E 1841-045 is well fitted with the standard power law plus blackbody (reduced
$\chi^2/dof \simeq 1.11$), nevertheless the two blackbody model gives also a rather good fit (reduced $\chi^2/dof \simeq
1.12$). Interestingly enough the blackbody parameters:
\begin{equation}
\label{black-bodies-1841}
\begin{split}
 &  R_1  \; =  \; 5.7 \; \; ^{+0.6}_{-0.5} \; \; \; \; Km \; \; \; , \; \; \;   T_1  \; =  \; 0.47 \; \pm \; 0.02  \; \;
             KeV \; \; \; , \\
 &  R_2  \; =  \; 0.36 \; ^{+0.08}_{-0.07} \; \; Km \; \; \; , \; \; \;   T_2  \; =  \; 1.5 \; ^{+0.2}_{-0.1}  \; \;  KeV
 \; \; \; ,
\end{split}
\end{equation}
are in agreement with  Eq.~(\ref{black-bodies}). Moreover, assuming that the power law component in the standard
parametrization of quiescent spectra account for the hot blackbody component in our parametrization, we find that the
suggestion  $L_{\gamma1} \; \simeq \; L_{\gamma2}$ in Eq.~(\ref{black-bodies}) is in agreement  with observations
(\cite{Marsden:2001}). Another interesting consequence of the anisotropic distribution of the surface temperature due to
strong magnetic fields is that the thermal surface blackbody radiation will be modulated by the stellar rotation. As a
matter of fact, Ozel (2002) argued that the observed properties of anomalous $X$-ray pulsars can be accounted for by
magnetars with a single hot region. It is remarkable that our interpretation explains naturally the observed change in
pulse profile of SGR 1900+14 following the 1998 August 27 giant flare. It seem that our picture is in fair qualitative
agreement with several observations. However, any further discussion of this matter goes beyond the aim of the present
paper.
\subsection{\normalsize{BURSTS}}
\label{bursts}
In the present Section we discuss how glitches in our magnetars give rise to the  burst activity from soft gamma-ray
repeaters and anomalous $X$-ray pulsars. We said in Sect.~\ref{glitches} that the energy released during a burst in a
magnetar is given by the magnetic energy directly injected into the magnetosphere, Eq.~(\ref{var-mag-ener-ext}). Before
addressing the problem of the dissipation of this magnetic energy in the magnetosphere, let us discuss what are the
observational signatures at the onset of the burst. Observations indicate that at the onset of giant bursts the flux
displays  a spike with a very short rise time $t_1$ followed by a rapid but more gradual decay time $t_2$. According to
our previous discussion, the onset of bursts is due to the positive variation of the surface magnetic field $\delta B_S$,
which in turn implies an sudden increase of the magnetic energy stored in the magnetosphere. Now, according to
Eq.~(\ref{mag-ener-ext}) we see that almost all the magnetic energy is stored in the region:
\begin{equation}
\label{4.54}
 R \;  \lesssim \; r  \;  \lesssim \; 10 \; R  \; \;  \; .
\end{equation}
So that the rise time is essentially the time needed to propagate in the magnetosphere the information that the surface
magnetic field is varied. Then we are lead to:

\begin{equation}
\label{4.55}
  t_1 \; \simeq \; 9 \; R  \; \simeq \; 3 \; 10^{-4}  \; sec  \; \; ,
\end{equation}
which indeed is in agreement with observations. On the other hand, in our proposal the decay time $t_2$ depends on the
physical properties of the magnetosphere. It is natural to identify $t_2$ with the time needed to the system to react to
the sudden variation of the magnetic field. In other words, we may consider the magnetosphere as a huge electric circuit
which is subject to a sudden increase of power from some external device. The electric circuit reacts to the external
injection of energy within  a transient time. So that, in our case the time $t_2$ is a function of the geometry and the
conducting properties of the magnetosphere. In general, it is natural to expect that $t_1 \, \ll \, t_2$ and the time
extension of the initial spike is:
\begin{equation}
\label{4.56}
 \delta \; t_{spike} \;  \simeq \; t_2 \; - \; t_1 \; \simeq \;  t_2  \;  \; \; .
\end{equation}
Remarkably, observations shows that the observed giant bursts are characterized by almost the same $\delta \; t_{spike}$:
\begin{equation}
\label{4.57}
 \delta \; t_{spike} \;  \simeq  \;  t_2  \; \simeq \; 0.1  \; sec  \; \; ,
\end{equation}
signalling that the structure of the magnetosphere of soft gamma-ray repeaters and anomalous $X$-ray pulsars are very
similar. Since the magnetic field is varied by  $\delta B_S$ in a time $\delta  t_{spike}$, then from Maxwell equations it
follows that it must be an induced electric field. To see this, let us consider the dipolar magnetic field in polar
coordinate:
\begin{equation}
\label{mag-polar}
\begin{split}
 B_r & = \; - \frac{2 \; B_S \; R^3 \; \cos \theta}{r^3}  \; \;  ,  \\
 B_\theta & = \; - \frac{ B_S \; R^3 \; \sin \theta}{r^3}  \; \;  ,  \\
B _\varphi & =  \; 0 \; ,
\end{split}
\end{equation}
Thus, observing that $\frac{\delta B_S}{\delta t_{spike}}$ is the time derivative of the magnetic field it is easy to find
the induced azimuthal electric field:
\begin{equation}
\label{electric}
E_\varphi  =  + \frac{\delta B_S}{\delta t_{spike}} \frac{ R^3  \; \sin \theta}{r^2}  \; \; \; , \; \; \; r \; \geq \; R
\; \; \; .
\end{equation}
To discuss the physical effects of the induced azimuthal electric field Eq.~(\ref{electric}), it is convenient to work in
the co-rotating frame of the star. We assume that the magnetosphere contains a neutral plasma. Thus, we see that charges
are suddenly accelerated by the huge induced azimuthal electric field $E_\varphi$ and thereby acquire an azimuthal
velocity $v_\varphi \simeq 1$ which is directed along the electric field for positive charges and in the opposite
direction for negative charges. Now, it is well known that relativistic charged particles moving in the magnetic field
${\bf{B}}({\bf{r}})$, Eq.~(\ref{mag-polar}), will emit synchrotron radiation (see for instance
\cite{wallace:1977,ginzburg:1979}). As we discuss below, these processes are able to completely dissipate the whole
magnetic energy injected into the magnetosphere following a glitch. However, before discuss this last point in details, we
would like to point out some general consequences which lead to well defined observational features. As we said before,
charges are accelerated by the electric field $E_\varphi$ thereby acquiring a relativistic azimuthal velocity. As a
consequence, they are subject to the drift Lorentz force ${\bf{F}} \, = \, q \, {\bf{v}}_\varphi \, \times \,
{\bf{B}}({\bf{r}})$, whose radial component is:
\begin{equation}
\label{F_r}
 F_r  \; = \;  -  q \, v_\varphi  B_\theta \;
   \simeq \;  + q \, v_\varphi B_S
  \sin \theta  \left ( \frac{ R}{r} \right )^3  ,
\end{equation}
while the $\theta$ component is:
\begin{equation}
\label{F_theta}
F_\theta \;  =  \; + q \, v_\varphi  B_r \;  \simeq \; - \, 2 \, q \,  v_\varphi
 B_S  \cos \theta  \left ( \frac{R}{r} \right )^3  .
\end{equation}
The radial component $F_r$ pushes both positive and negative charges radially outward. Then, we see that the plasma in the
outermost part of the magnetosphere is subject to a intense radial centrifugal force, so that the plasma must flow
radially outward giving rise to a blast wave. On the other hand,  $F_\theta$ is centripetal in the upper hemisphere and
centrifugal in the lower hemisphere. As a consequence, in the lower hemisphere charges are pushed towards the magnetic
equatorial plane $\cos \theta = 0$, while in the upper hemisphere (the north magnetic pole) the centripetal force gives
rise to a rather narrow jet along the magnetic axis. As a consequence, at the onset of the giant burst there is an almost
spherically symmetric outflow from the pulsar together with a collimated jet from the north magnetic pole. Interestingly
enough, a fading radio source has been seen from SGR 1900+14 following the  August 27 1998 giant flare
(\cite{Frail:1999}). The radio afterglow is consistent with an outflow expanding subrelativistically into the surrounding
medium. This is in agreement with our model once one takes into account that the azimuthal electric field is rapidly
decreasing with the distance from the star, so that $v_\varphi \lesssim 1$ for the plasma in the outer region of the
magnetosphere. However, we believe that the most compelling evidence in favour of our proposal comes from the detected
radio afterglow following the 27 December 2004 gigantic flare from
SGR 1806-20 (\cite{Gaensler:2005,Cameron:2005,Wang:2005,Granot:2005,Gelfand:2005}). Indeed, the fading radio source from
{\it {SGR 1806-20}} has similar properties as that observed from {\it {SGR 1900+14}}, but much higher energy. The
interesting aspect is that in this case the spectra of the radio afterglow showed clearly the presence of the expected
spherical non relativistic expansion together with a sideways expansion of a jetted explosion (see Fig.~1 in
\cite{Gaensler:2005} and Fig.~1 in \cite{Cameron:2005}).  Finally, the lower limit of the outflow $E \gtrsim 10^{44.5} \;
ergs$ (\cite{Gelfand:2005}) implies that the blast wave and the jet dissipate only a small fraction of the burst energy
which is about $10^{46} \; ergs$ (see Section~\ref{braking}). Thus, we infer that almost all the burst energy must be
dissipated into the magnetosphere. In the co-rotating frame of the star the plasma, at rest before the onset of the burst,
is suddenly accelerated by the induced electric field thereby acquiring an azimuthal velocity $v_\varphi \simeq 1$. Now,
relativistic charges are moving in the dipolar magnetic field of the pulsar. So that, they will lose energy by emitting
synchrotron radiation until they come at rest. Of course, this process, which involves charges that are distributed in the
whole magnetosphere, will last for a time $t_{dis}$ much longer that $\delta t_{spike}$. Actually, $t_{dis}$ will depend
on the injected energy, the plasma distribution and the magnetic field strength. Moreover, one should also take care of
repeated charge and photon scatterings. So that it is not easy to estimate $t_{dis}$ without a precise knowledge of the
pulsar magnetosphere. At the same time, the fading of the luminosity with time, the so-called light curve $L(t)$, cannot
be determined without a precise knowledge of the microscopic dissipation mechanisms. However, since the dissipation of the
magnetic energy involves the whole magnetosphere, it turns out that we may  accurately reproduce the time variation of
$L(t)$ without a precise knowledge of the microscopic dissipative mechanisms. Indeed, in Sect.~\ref{light} we develop an
effective description where our ignorance on the microscopic dissipative processes is encoded in few macroscopic
parameters, which allows us to determine the light curves. In the remaining of the present Section we investigate the
spectral properties of the luminosity. To this end, we need to consider the synchrotron radiation spectral distribution.
Since radiation from electrons is far more important than from protons, in the following we shall focus on electrons. It
is well known that the synchrotron radiation will be mainly at the frequencies (\cite{wallace:1977,ginzburg:1979}):
\begin{equation}
\label{omega_m}
\omega_m \; \simeq \; \gamma^2 \; \frac{e  B}{ m_e} \; ,
\end{equation}
where $\gamma$ is the electron Lorentz factor. Using Eq.~(\ref{mag-polar}) we get:
\begin{equation}
\label{cyclo}
 \omega(r)  \; \simeq \; \gamma^2 \;   \frac{e  B_S}{ m_e} \;  \left ( \frac{ R}{r} \right )^3 \; \; , \; \;
                R \;  \lesssim \; r  \;  \lesssim \; 10 \; R  \; \;  \; .
\end{equation}
It is useful to numerically estimate the involved frequencies. To this end, we consider the giant flare of 1998, August 27
from SGR 1900+14:
\begin{equation}
\label{1900-par}
 B_S \; \simeq \; 7.4 \; 10^{14} \; Gauss \; \; \; , \; \; \; \frac{\delta B_S}{B_S} \; \simeq \; 10^{-2}  \; \;  \; .
\end{equation}
So that, from Eq.~(\ref{cyclo}) it follows:
\begin{equation}
\label{4.62}
  \omega(r)  \; \simeq \; \gamma^2 \; 8.67 \; MeV \;  \left ( \frac{ R}{r} \right )^3 \; \; , \; \;
                R \;  \lesssim \; r  \;  \lesssim \; 10 \; R  \; \;  \; ,
\end{equation}
or
\begin{equation}
\label{4.63}
  \omega_1  \, \simeq \, \gamma^2 \, 8.67 \; KeV  \; \lesssim \; \; \omega \; \; \lesssim \;
        \omega_2  \, \simeq \, \gamma^2 \, 8.67 \; MeV   \; \;  \; .
\end{equation}
The power injected into the magnetosphere is supplied by the azimuthal electric field during the initial hard spike. So
that to estimate the total power we need to evaluate the power supplied by the azimuthal electric field. Let us consider
the infinitesimal volume $dV=r^2 \sin \theta dr d\theta d\varphi$; the power supplied by the induced electric field
$E_\varphi$ in $dV$:
\begin{equation}
\label{inf-pow}
 d \dot{W}_{E_\varphi} \; \simeq \;  n_e \; e \; \frac{\delta B_S}{\delta t_{spike}}  \; v_\varphi \;  R^3  \;
 \sin^2 \theta dr d\theta d\varphi  \; ,
\end{equation}
where $n_e$ is  the electron number density. Since the magnetosphere is axially symmetric it follows that $n_e$ cannot
depend on $\varphi$. Moreover, within our theoretical uncertainties we may neglect the dependence on $\theta$. So that,
integrating over $\theta$ and $\varphi$ we get:
\begin{equation}
\label{4.68}
 d \dot{W}_{E_\varphi} \; \simeq \; 2 \; \pi^2 \;  n_e \; e \; \frac{\delta B_S}{\delta t_{spike}}  \; v_\varphi \;  R^3  \;
 dr   \; .
\end{equation}
In order to determine the spectral distribution of the supplied power, we note that to a good approximation all the
synchrotron radiation is emitted at $\omega_m$, Eq.~(\ref{omega_m}). So that, we may use Eq.~(\ref{4.62}) to get:
\begin{equation}
\label{dr-dnu}
 - \; dr \;  \simeq \; \frac{R}{3} \;  \gamma^{\frac{2}{3}} \; \left ( \frac{e  B_S}{m_e}  \right )^{1/3}
 \frac{1}{\omega^{\frac{4}{3}}} \; d\omega \; \; .
\end{equation}
Inserting Eq.~(\ref{dr-dnu}) into Eq.~(\ref{4.68}) we obtain the  spectral power:
\begin{equation}
\label{spect-pow}
 F(\omega) \; d\omega \; \simeq \;  \frac{2 \pi^2}{3} \; n_e \; e \; \frac{\delta B_S}{\delta t_{spike}}  \; v_\varphi \;  R^4  \;
  \; \gamma^{\frac{2}{3}} \; \left ( \frac{e  B_S}{m_e}  \right )^{1/3}
 \frac{1}{\omega^{\frac{4}{3}}} \; d\omega \; ,
\end{equation}
while the total luminosity is given by:
\begin{equation}
\label{lum-tot}
L  \; = \; \int_{\omega_1}^{\omega_2} \; F(\omega) \; d\omega \; \; .
\end{equation}
Note that $L$ is the total luminosity injected into the magnetosphere during the initial hard spike. So that, since the
spike lasts $\delta \; t_{spike}$, we have:
\begin{equation}
\label{4.72}
E_{\emph{burst}}   \; \simeq  \; \delta \; t_{spike} \; L  \; \; ,
\end{equation}
where $E_{\emph{burst}}$ is the total burst energy. In the case of the 1998 August 27 giant burst from SGR 1900+14 the
burst energy is given by Eq.~(\ref{GB-1900}). Thus, using Eqs.~(\ref{4.72}) and ~(\ref{4.57}) we have:
\begin{equation}
\label{4.73}
L  \; \simeq \; 10^{45}  \; \frac{ergs}{sec} \; \; \; ,
\end{equation}
which, indeed, is in agreement with observations. It is worthwhile to estimate the electron number density needed to
dissipate the magnetic energy injected in the magnetosphere. To this end, we assume an uniform number density. Thus, using
Eqs.~(\ref{lum-tot}), (\ref{spect-pow}) and (\ref{4.63}) we get:
\begin{equation}
\label{4.74}
L  \; \simeq \; 18 \; \pi^2 \; n_e \; e \; \frac{\delta B_S}{\delta t_{spike}} \;  R^4   \; \; ,
\end{equation}
where we used $v_\varphi \simeq 1$. Specializing to the  August 27 giant burst we find:
\begin{equation}
\label{4.75}
 n_e \; \simeq \; 2.0  \; 10^{14}  \; cm^{-3}  \;  \; \; ,
\end{equation}
indeed quite a reasonable value. Soon after the initial spike, the induced azimuthal electric field vanishes and the
luminosity decreases due to dissipative processes in the magnetosphere. As thoroughly discussed in Sect.~\ref{light}, it
is remarkable that the fading of the luminosity can be accurately reproduced without a precise knowledge of the
microscopic dissipative mechanisms. So that combining the time evolution of the luminosity $L(t)$, discussed in
Sect.~\ref{light}, with the spectral decomposition we may obtain the time evolution of the spectral components. In
particular, firstly we show that starting from Eq.~(\ref{spect-pow}) the spectral luminosities can be accounted for by two
blackbodies and a power law. After that, we discuss the time evolution of the three different spectral components. \\
The spectral decomposition Eq.~(\ref{spect-pow}) seems to indicate that the synchrotron radiation follows a power law
distribution. However,  one should take care of reprocessing effects which redistribute the spectral distribution. To see
this, we note that photons with energy $\omega \geq 2 \, m_e$ quickly will  produce copiously almost relativistic
$e^{\pm}$ pairs. Now, following Duncan \& Thompson (1995), even if the particles are injected steadily in a time $\delta
t_{spike}$, it is easy to argue that the energy of relativistic particles is rapidly converted due to comptonization to
thermal photon-pair plasma. Since the pair production is quite close to the stellar surface, we may adopt the rather crude
approximation of an uniform magnetic filed $B \simeq B_S$ throughout the volume $V_{plasma} \simeq 12 \, \pi \, R^3$
(Duncan \& Thompson 1995). Since typical magnetic fields in magnetars are well above $B_{QED}$, electrons and positrons
sit in the lowest Landau levels. In this approximation we deal with an almost one dimensional pair plasma whose energy
density is (\cite{Duncan:1995}):
\begin{equation}
\label{4.76}
 u_e \; \simeq \; m_e \; (n_{e^+} \; + n_{e^-}) \; \simeq \frac{2}{(2 \pi)^{\frac{3}{2}}}  \; e B_S \;  m_e^2 \;
         \left ( \frac{T_{plasma}}{m_e} \right )^{\frac{1}{2}} \exp (-  \frac{m_e}{T_{plasma}})  \;  \; \; ,
\end{equation}
for $T_{plasma} < m_e$, $T_{plasma}$ being the plasma temperature. Thus, the total energy density of the thermal
photon-pair plasma is:
\begin{equation}
\label{4.77}
 u \; = \; u_e \; + \; u_{\gamma} \; \simeq \;  u_e \; + \; \frac{\pi^2}{15} \; T_{plasma}^4 \; \; \; .
\end{equation}
The plasma temperature is determined by equaling the thermal energy Eq.~(\ref{4.77}) with the fraction of burst energy
released in the spectral region $\omega \geq 2 \, m_e$. It is easy to find:
\begin{equation}
\label{4.78}
 E_{pairs} \; \simeq \; 0.147 \; E_{burst} \; \; \; ,
\end{equation}
where for the numerical estimate we approximated $\omega_1 \simeq 10 \, KeV$ and $\omega_2 \simeq 10 \, MeV$,
corresponding to mildly relativistic electrons in the magnetosphere. So that we have:
\begin{equation}
\label{4.79}
 \frac{2}{(2 \pi)^{\frac{3}{2}}}  \; e B_S \;  m_e^2 \;
         \left ( \frac{T_{plasma}}{m_e} \right )^{\frac{1}{2}} \exp (-  \frac{m_e}{T_{plasma}})  \;  + \;
 \; \frac{\pi^2}{15} \; T_{plasma}^4 \; \simeq \; \frac{ E_{pairs}}{V_{plasma}} \; \; \; .
\end{equation}
In the case of August 27 giant burst from SGR 1900+14 we find:
\begin{equation}
\label{4.80}
 \sqrt{x} \;  \exp (-  \frac{1}{x}) \; +  \; 0.311 \; x^4  \simeq \; 1.32 \; 10^{-2} \; \; , \; \;
 x \; = \; \frac{T_{plasma}}{m_e} \; \; \; ,
\end{equation}
whose solution gives $T_{plasma} \simeq 135 \, KeV$. However, this is not the end of the whole story. Indeed, our thermal
photon-pair plasma at temperature  $T_{plasma}$ will be reprocessed by thermal electrons on the surface which are at
temperature of the thermal quiescent emission $T_Q  \lesssim 1 \, KeV$. So that, photons at temperature $T_{plasma} \gg
T_Q$ are rapidly cooled by Thompson scattering off electrons in the stellar atmosphere, which extends over several
hundreds \emph{fermis} beyond the edge of the star. The rate of change of the radiation energy density is given by (for
instance, see \cite{Peebles:1993}):
\begin{equation}
\label{4.81}
 \frac{1}{u_{\gamma}} \; \frac{d u_{\gamma}}{d t} \;  \simeq \; \frac{4 \, \sigma_T \, n_Q}{m_e} \;
 ( T_Q - T_{plasma} ) \; \; \; ,
\end{equation}
where $n_Q$ is the number density of electrons in the stellar atmosphere. The electron number density in the atmosphere of
a p-star is of the same order as in strange stars, where $n_Q \simeq 10^{33} \, cm^{-3}$ (\cite{alcock:1986}). So that,
due to the very high electron density of electrons near the surface of the star, the thermal photon-pair plasma is
efficiently cooled to a final temperature much smaller than $T_{plasma}$. At the same time, the energy transferred to the
stellar surface leads to an increase of the effective quiescent temperature. Therefore we are lead to conclude that during
the burst activity the quiescent luminosity must increase. Let $T_1$ be the final plasma temperature, then we see that the
thermal photon-pair plasma contribution to the luminosity can be accounted for with an effective blackbody with
temperature $T_1$ and radius $R_1$ of the order of the stellar radius. As a consequence the resulting blackbody luminosity
is:
\begin{equation}
\label{4.82}
 L_{1} \; = \; 4 \, \pi \, R_1^2 \, \sigma_{SB} \, T_1^4 \; \; \; , \; \; \; R_1 \; \lesssim \; R \; \; .
\end{equation}
In general, the estimate of the effective blackbody temperature $T_1$ is quite difficult. However, according to
Eq.~(\ref{4.78}) we known that $L_1$ must account for about  $0.147$ of the total luminosity. So that we have:
\begin{equation}
\label{4.83}
 L_{1}(t) \; \simeq \; 0.147 \; L(t) \; \; .
\end{equation}
This last equation allow us to determine the blackbody temperature. For instance, soon after the hard spike we have $L(0)
\simeq \frac{E_{\emph{burst}}}{\delta  t_{spike}} \simeq 10^{45}  \; \frac{ergs}{sec}$ for the giant burst from SGR
1900+14. Thus, using $R_1  \simeq  R$, from Eq.~(\ref{4.83}) we get:
\begin{equation}
\label{4.84}
 T_1(0) \;  \simeq \; 61 \; KeV \; \; \; ,
\end{equation}
with surface luminosity $L_1(0) \simeq  10^{44}  \; \frac{ergs}{sec}$. \\
Let us consider the remaining spectral power with $\omega \lesssim 2 m_e$. We recall that the spectral power
Eq.~(\ref{spect-pow}) originates from the power supplied by the induced electric field Eq.~(\ref{4.68}). It is evident
from Eq.~(\ref{4.68}) that, as long as $v_\varphi \simeq 1$, the power supplied by the electric field $E_\varphi$ does not
depend on the mass of accelerated charges. Since the plasma in the magnetosphere is neutral, it follows that protons
acquire the same energy as electrons. On the other hand, since the protons synchrotron frequencies are reduced by a factor
$\frac{m_e}{m_p}$, the protons will emit synchrotron radiation near $\omega_1$. As a consequence, photons with frequencies
near $\omega_1$ suffer resonant synchrotron scattering, which considerably redistribute the available energy over active
modes. On the other hand, for $\omega \gg \omega_1$ the spectral power will follows the power law Eq.~(\ref{spect-pow}).
Thus, we may write:
\begin{equation}
\label{4.85}
 F(\omega) \;  \sim \;  \frac{1}{\omega^{\frac{4}{3}}} \; \; \; \; \; , \; \; \; \; \; 5 \,
            \omega_1 \; \lesssim \; \omega \; \lesssim \;  2 \, m_e \; \; \; ,
\end{equation}
where we have somewhat arbitrarily assumed the low energy cutoff $\thicksim 5 \, \omega_1$. On the other hand, for $\omega
\; \lesssim \; 5 \, \omega_1$ the redistribution of the energy by resonant synchrotron scattering over electron and proton
modes lead to an effective description of the relevant luminosity as thermal blackbody with effective temperature and
radius $T_2$ and $R_2$, respectively. Obviously, the blackbody radius $R_2$ is fixed by the geometrical constrain that the
radiation is emitted in the magnetosphere at distances $r \lesssim 10 \, R$. So that we have:
\begin{equation}
\label{4.86}
 R_2  \;  \lesssim  \; 10 \; R \; \;  \; .
\end{equation}
The effective blackbody temperature $T_2$ can be estimate by observing that the  integral of the spectral power up to $5
\, \omega_1$ account for about the $60 \; \% $ of the total luminosity. Thus, we have:
\begin{equation}
\label{4.87}
 L_{2}(t) \; \simeq \; 0.60 \; L(t) \; \; ,
\end{equation}
where
\begin{equation}
\label{4.88}
 L_{2} \; = \; 4 \, \pi \, R_2^2 \, \sigma_{SB} \, T_2^4 \; \; \; , \; \; \; R_2 \; \lesssim \; 10 \; R \; \; .
\end{equation}
Equations~(\ref{4.87}) and (\ref{4.88}) can be used to  to determine the effective blackbody temperature. If we consider
again  the giant burst from SGR 1900+14, soon after the hard spike, assuming  $R_2 \simeq 10 \, R$, we readily obtain:
\begin{equation}
\label{4.89}
 T_2(0) \;  \simeq \; 27 \; KeV \; \; \; .
\end{equation}
To summarize, we have found that  the spectral luminosities can be accounted for by two blackbodies and a power law. In
particular for the blackbody components we have:
\begin{equation}
\label{4.90}
\begin{split}
 &  R_1  \; \lesssim \; R  \; \; , \; \; R_2 \lesssim \; 10 \; R \; \; ; \; \;
    T_2  \;  <  \; T_1 \; \; , \; \;  R_1  \; <  \; R_2 \;  \; \\
 &  \; \; \; \;  \; \; \; \; L_{1} \; \simeq \; 0.15 \; L \; \; \;, \; \; \;
    L_{2} \; \simeq \; 0.60 \; L \; \; \; .
\end{split}
\end{equation}
Interestingly enough, Eq.~(\ref{4.90}) displays an anticorrelation between blackbody radii and temperatures, in fair
agreement with observations. Moreover, the remaining $25 \%$ of the total luminosity is accounted for by a power law
leading to the high energy tail of the spectral flux:
\begin{equation}
\label{4.91}
\frac{d N}{d E}  \;  \sim \;   E^{- \alpha}  \; \; \; , \; \;  \; \alpha \; \simeq \; 2.33 \; \; \; \; ,
\end{equation}
extending up to $E \simeq 2 m_e \simeq 1 \, MeV$. Indeed, the high energy power law tail is clearly displayed in the giant
flare from SGR 1900+14 (see Fig.~3 in \cite{Feroci:1999}), and in the recent gigantic flare from SGR
1806-20 (see Fig.~4 in \cite{Hurley:2005}). \\
It is customary to fit the spectra with the sum of a power law and an optically thin thermal bremsstrahlung. It should be
stressed that the optically thin thermal bremsstrahlung model is purely phenomenological. In view of this, a direct
comparison of our proposal with data is problematic. Fortunately, Feroci et al. (2001) tested several spectral functions
to the observed spectrum in the afterglow of the giant outburst from SGR 1900+14. In particular they found that, in the
time interval $68 \, sec \, \lesssim \, t \, \lesssim \, 195 \, sec$, the minimum $\chi^2$ spectral model were composed by
two blackbody laws plus a power law. By fitting the time averaged spectra they reported(~\cite{Feroci:2001}):
\begin{equation}
\label{4.92}
 T_2 \;  \simeq \; 9.3 \; KeV \; \;  , \; \; T_1 \;  \simeq \; 20.3 \; KeV \; \; \; , \; \; \; \alpha \; \simeq \; 2.8 \; \; .
\end{equation}
Moreover, it turns out that the power law accounts for approximately $10 \%$ of the total energy above $25 \; KeV$, while
the low temperature blackbody component accounts for about  $85 \%$ of the total energy above $25 \; KeV$. In view of our
neglecting  the contribution to energy from protons, we see that our proposal is in accordance with the observed energy
balance. Unfortunately, in Feroci et al. (2001) the blackbody radii are not reported. To compare our estimate of the
blackbody temperatures with the fitted values in Eq.~(\ref{4.92}), we note that our values reported in Eqs.~(\ref{4.84})
and (\ref{4.89}) correspond to the blackbody temperatures soon after the initial hard spike. Thus, we need to determine
how the blackbody temperatures evolve with time. To this end, we already argued that  soon after the initial spike the
luminosity decreases due to dissipative processes in the magnetosphere. In Sect.~\ref{light} we show that the fading of
the luminosity can be accurately reproduced without a precise knowledge of the microscopic dissipative mechanisms. In
particular, the relevant light curve is given by Eqs.~(\ref{5.6}) and (\ref{5.10}). At $t=0$ we have seen that the total
luminosity is well described by three different spectral components. During the fading of the luminosity, it could happens
that microscopic dissipative processes modify the different spectral components. However, it is easy to argue that this
does not happens. The crucial point is that the three spectral components originate from emission by a macroscopic part of
the magnetosphere; moreover the time needed to modify a large volume of magnetosphere by microscopic processes is much
larger than the dissipation time $t_{dis} \, \sim \, 10^2 \, sec$. Then we conclude that, even during the fading of the
luminosity, the decomposition of the luminosity into three different spectral components retain its validity. Now, using
the results in Sect.~\ref{light}, we find:
\begin{equation}
\label{4.93}
 \frac{L(t \, \simeq \, 68 \, sec)}{L(0)}  \; \simeq \; 3.67 \; 10^{-2} \; \; \; , \; \; \;
\frac{L(t \, \simeq \, 195 \, sec)}{L(0)}  \; \simeq \; 1.67 \; 10^{-2} \; \; \; .
\end{equation}
Combining Eq.~(\ref{4.93}) with Eqs.~(\ref{4.82}), (\ref{4.83}), (\ref{4.87}) and (\ref{4.88}) we obtain:
\begin{equation}
\label{4.94}
\begin{split}
 &  T_2(t \, \simeq \, 68 \, sec) \;  \simeq \; 11.8 \; KeV \; \; , \; \; T_2(t \, \simeq \, 195 \, sec)
        \;  \simeq \; 9.7 \; KeV  \; \; \; ,  \\
 &  T_1(t \, \simeq \, 68 \, sec) \;  \simeq \; 26.7 \; KeV \; \; , \; \; T_1(t \, \simeq \, 195 \, sec)\;  \simeq \; 21.9 \; KeV
     \; ,
\end{split}
\end{equation}
in reasonable agreement with Eq.~(\ref{4.92}). Finally, let us comment on the time evolution of the spectral exponent
$\alpha$ in the power law Eq.~(\ref{4.91}). From Eq.~(\ref{spect-pow}) it follows that high energy modes have less energy
to dissipate. Accordingly, once a finite amount of energy is stored into the magnetosphere, the modes with higher energy
become inactive before the lower energy modes. As a consequence, the effective spectral exponent will increase with time
and the high energy tail of the emission spectrum becomes softer. This explains also why the fitted spectral exponent
$\alpha$ in Eq.~(\ref{4.92}) is slightly higher than our estimate in Eq.~(\ref{4.91}).
\subsection{\normalsize{HARDNESS RATIO}}
\label{hardness}
\begin{figure}[ht]
   \centering
\includegraphics[width=0.90\textwidth,clip]{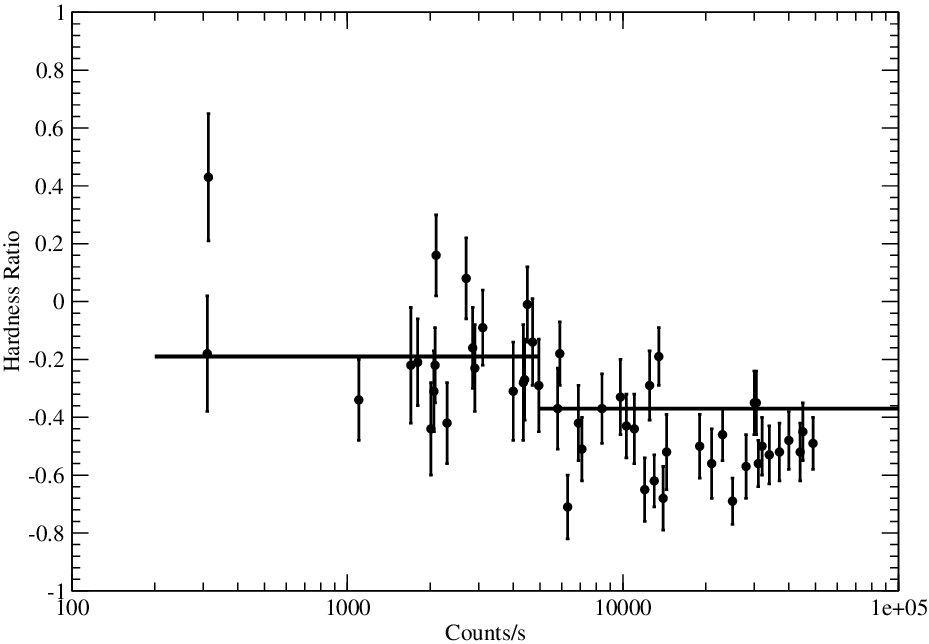}
\caption{\label{fig_4}
Hardness-intensity plot of the time resolved hardness ratio, Eq.~(\ref{4.95}). Data have been extract from Fig.~3 in Gotz
et al. (2004). Full lines are our Eqs.~(\ref{4.100}) and (\ref{4.102}).}
\end{figure}
Recently, it has been reported evidence for a hardness-intensity anti correlation within bursts from SGR 1806-20
(\cite{Gotz:2004}). Indeed, Gotz et al. (2004) reported observations of the soft gamma ray repeaters SGR 1806-20 obtained
in October 2003, during a period of bursting activity. They found that some bursts showed a significant spectral
evolution. However, in the present Section we focus on the remarkable correlation between hardness ratio and count rate.
Following  Gotz et al. (2004) we define the hardness ratio as:
\begin{equation}
\label{4.95}
 HR \;  = \frac{H \; - \; S}{H \; + \; S} \; \; \; ,
\end{equation}
where $H$ and $S$ are the background subtracted counts in the ranges $40-100 \; KeV$ and $15-40 \; KeV$ respectively. In
Figure~\ref{fig_4} we report the hardness ratio data extracted from Fig.~3 in  Gotz et al. (2004). \\
A few comments are in  order. First, the hardness ratio becomes negative for large enough burst intensities. Moreover,
there is  a clear decrease of the hardness ratio with increasing burst intensities. Now we show that within our approach
we may explain why the hardness ratio is negative and decreases with increasing burst intensities. To see this, we note
that the hardness ratio Eq.~(\ref{4.95}) is defined in terms of total luminosities in the relevant spectral intervals.
Thus, to determine the total luminosity in the spectral interval $\omega_1-\omega_2$ we may use:
\begin{equation}
\label{4.96}
L (\omega_1-\omega_2) \; = \; \int_{\omega_1}^{\omega_2} \; F(\omega) \; d\omega \; \; ,
\end{equation}
where $ F(\omega)$ is given by Eq.~(\ref{spect-pow}). A straightforward integration gives:
\begin{equation}
\label{4.97}
L(\omega_1-\omega_2) \; \simeq \; 2 \; \pi^2 \; n_e \; e \; \frac{\delta B_S}{\delta t_{spike}} \;  R^4 \;
                      \left ( \frac{r_1}{R} \; - \; \frac{r_2}{R} \right ) \; \; ,
\end{equation}
where $r_1$ and $r_2$ are given by:
\begin{equation}
\label{4.98}
  \omega_{1,2}  \; \simeq \; \gamma^2 \; \frac{e  B_S}{ m_e} \; \left ( \frac{ R}{r_{1,2}} \right )^3 \; \; \; .
\end{equation}
Assuming $\gamma \thicksim 1$, we may rewrite Eq.~(\ref{4.98}) as:
\begin{equation}
\label{4.99}
  \omega_{1,2}  \; \simeq \; 10 \; MeV \;  \left ( \frac{ R}{r_{1,2}} \right )^3 \; \; \; .
\end{equation}
Using Eqs.~(\ref{4.97}) and ~(\ref{4.99}) it is easy to determine the hardness ratio:
\begin{equation}
\label{4.100}
 HR \;  = \frac{L(40-100 \; KeV) \; - \; L(15-40 \; KeV)}{L(40-100 \; KeV) \; + \; L(15-40 \; KeV)} \; \simeq \;
          \frac{1.66 \; - \; 2.44}{ 1.66 \; + \; 2.44} \; \simeq \; - \; 0.19 \; \; .
\end{equation}
In Figure~\ref{fig_4} we display our estimate of the hardness ratio Eq.~(\ref{4.100}). We see that data are in quite good
agreement with Eq.~(\ref{4.100}) at least up to count rate $\thicksim \, 5 \, 10^3 \; counts/sec$. For larger count rates
data seem to lie below our value. We believe that, within our approach, there is a natural explanation for this effect.
Indeed, for increasing count rates we expect that the hard tail $\omega \gtrsim 2 \, m_e \simeq 1 \, MeV$ of the spectrum
will begin to contribute to the luminosity. According to the discussion in Sect.~\ref{bursts} these hard photons are
reprocessed leading to an effective blackbody with temperature $T_1$. Now, for small and intermediate bursts the blackbody
temperature $T_1$ is considerably smaller than Eq.~(\ref{4.84}), so that the effective blackbody contributes mainly to the
soft tail of the spectrum. Obviously, the total luminosity of the effective blackbody is:
\begin{equation}
\label{4.101}
L(1-10 \; MeV) \; \simeq \; 2 \; \pi^2 \; n_e \; e \; \frac{\delta B_S}{\delta t_{spike}} \;  R^4 \;
                      \left ( 2.15 \; - \; 1 \right ) \; \; .
\end{equation}
Since this luminosity contributes to the soft part of the emission spectrum, Eq.~(\ref{4.100}) gets modified as:
\begin{equation}
\label{4.102}
 HR \;  \simeq \;  \frac{1.66 \; - \; 3.59}{ 1.66 \; + \; 3.59} \; \simeq \; - \; 0.37 \; \; .
\end{equation}
Equation~(\ref{4.102}) is displayed in Fig.~\ref{fig_4} for rates $\gtrsim  5 \, 10^3 \; counts/sec$. Note that we did not
take into account the proton contribution to the luminosity. Observing that protons contribute mainly to luminosities at
low energy $\omega \lesssim 10 \,  KeV$, we see that adding the proton contributions leads to smaller hardness ratios
bringing our estimates to a better agreement with data. In any case, we see that our theory allows to explain in a natural
way the anti correlation between hardness ratio and intensity.
\section{\normalsize{LIGHT CURVES}}
\label{light}
In our magnetar theory the observed burst activities are triggered by glitches which inject magnetic energy into the
magnetosphere where, as discussed in previous Sections, it is dissipated. As a consequence the observed luminosity is time
depended. In this Section we set up an effective description which allows us to determine the light curves, i.e. the time
dependence of the luminosity. In general, the  energy injected into the magnetosphere after the glitch decreases due to
dissipative effects described in Sect.~\ref{bursts}, leading to the luminosity  $L(t) = - \frac{d E(t)}{dt}$. Actually,
the precise behavior of $L(t)$ is determined once the dissipation mechanisms are known. However, since the dissipation of
the magnetic energy involves the whole magnetosphere, we may accurately reproduce the time variation of $L(t)$ without a
precise knowledge of the microscopic dissipative mechanisms. Indeed, on general grounds we have that the dissipated energy
is given by:
\begin{equation}
\label{5.1}
L(t) \; \; = \; \; - \; \frac{d E(t)}{dt} \; \; = \; \kappa(t) \; E^\eta \; \; , \; \; \eta \; \leq \; 1 \; \; \; ,
\end{equation}
where $\eta$ is the efficiency coefficient. Obviously the parameter $\kappa(t)$ does depend on the physical parameters of
the magnetosphere. For an ideal system, where the initial injected energy is huge, the linear regime, where $\eta = 1$, is
appropiate. Moreover, we expect that the dissipation time $ \thicksim \frac{1}{\kappa}$ is much smaller than the
characteristic time needed to macroscopic modifications of the magnetosphere. Thus, we may safely assume $\kappa(t) \simeq
\kappa_0$ constant. So that we get:
\begin{equation}
\label{5.2}
L(t) \; \; = \; \; - \; \frac{d E(t)}{dt} \; \; \simeq \; \kappa_0 \; E \; \; .
\end{equation}
It is straightforward to solve Eq.~(\ref{5.2}):
\begin{equation}
\label{5.3}
 E(t)  \; = \; E_0 \; \exp(- \frac{t}{\tau_0}) \; \; \; \; , \; \; \;
 L(t) \;  = \;  L_0 \; \exp(- \frac{t}{\tau_0}) \; \; , \; \; L_0 \; = \; \frac{E_0}{\tau_0} \; \; , \; \;
 \tau_0 \; = \; \frac{1}{\kappa_0} \; \; \; .
\end{equation}
Note that the dissipation time $\tau_0  = \frac{1}{\kappa_0}$ encodes all the physical information on the microscopic
dissipative phenomena. Since the injected energy is finite, the dissipation of energy degrades with the decreasing of the
available energy. Thus, the relevant equation is Eq.~(\ref{5.1}) with $\eta < 1$. In this case, solving Eq.~(\ref{5.1}) we
find:
\begin{equation}
\label{5.4}
 L(t) \;  = \;  L_0 \; \left ( 1 - \frac{t}{t_{dis}} \right )^{\frac{\eta}{1-\eta}}  \; \; ,
\end{equation}
where we have introduced the dissipation time:
\begin{equation}
\label{5.5}
t_{dis}  \; \; = \; \; \frac{1}{\kappa_0} \; \frac{E_0^{1-\eta}}{1-\eta}  \; \; .
\end{equation}
Then, we see that the time evolution of the luminosity is linear up to some time $t_{break}$, after that we have a break
from the linear regime $\eta = 1$ to a non linear regime with $\eta < 1$. If we indicate with $t_{dis}$ the total
dissipation time,  we get:
\begin{equation}
\label{5.6}
\begin{split}
L(t) \; &    = \;  L_0 \; \exp(- \frac{t}{\tau_0}) \; \; \; \;  \; \; ,  \; \; \; \; \; \; \; \; \; \; \; \; \; \; \; \;
                   \; \; \; \; \; \; \; \; \; \; \; \; \; \; \; \;\; \; 0 \; < \; t \; < \; t_{break}  \; \;  ,  \\
L(t) \; &    = \;  L(t_{break}) \;  \left ( 1 - \frac{t-t_{break}}{t_{dis}-t_{break}} \right )^{\frac{\eta}{1-\eta}}  \;
\; ,
               \; \; \;  \; \; \; \;  \; \; t_{break}  \; < \; t \; < \; t_{dis}   \; \;  .
\end{split}
\end{equation}
Equation~(\ref{5.6}) is relevant to describe the light curve after a giant burst, where there is a huge amount of magnetic
energy dissipated into the magnetosphere. It is interesting to compare our light curves, Eq.~(\ref{5.6}), with the
standard magnetar model. The decay of the luminosity in the standard magnetar model is due to the evaporation by a
fireball formed after a giant burst and trapped onto the stellar surface (\cite{Duncan:1996,Duncan:2001}). Indeed, Feroci
et al. (2001) considered the light curves after the giant flare of 1998, August 27 from SGR 1900+14, and the giant flare
of 1979 March 5 from SGR 0526-66. Assuming that the luminosity varies as a power of the remaining fireball energy $L
\thicksim E^a$, they found:
\begin{equation}
\label{5.7}
 L(t) \;  = \;  L_0 \; \left ( 1 - \frac{t}{t_{evap}} \right )^{\frac{a}{1-a}}  \; \; ,
\end{equation}
where $t_{evap}$ is the time at which the fireball evaporates, and the index $a$ accounts for the geometry and the
temperature distribution of the trapped fireball. For a spherical fireball of uniform temperature $a = \frac{2}{3}$, so
that the index $a$ must satisfies the constrain:
\begin{equation}
\label{5.8}
a \;  \leqslant \frac{2}{3}  \;   \; \; .
\end{equation}
Note that our Eq.~(\ref{5.6}) reduces to Eq.~(\ref{5.7}) if $t_{break} = 0$ and $\eta = a$. However, we stress that our
efficiency exponent  must satisfy the milder constraint $\eta \leqslant 1$.

\begin{figure}[ht]
   \centering
\includegraphics[width=0.9\textwidth,clip]{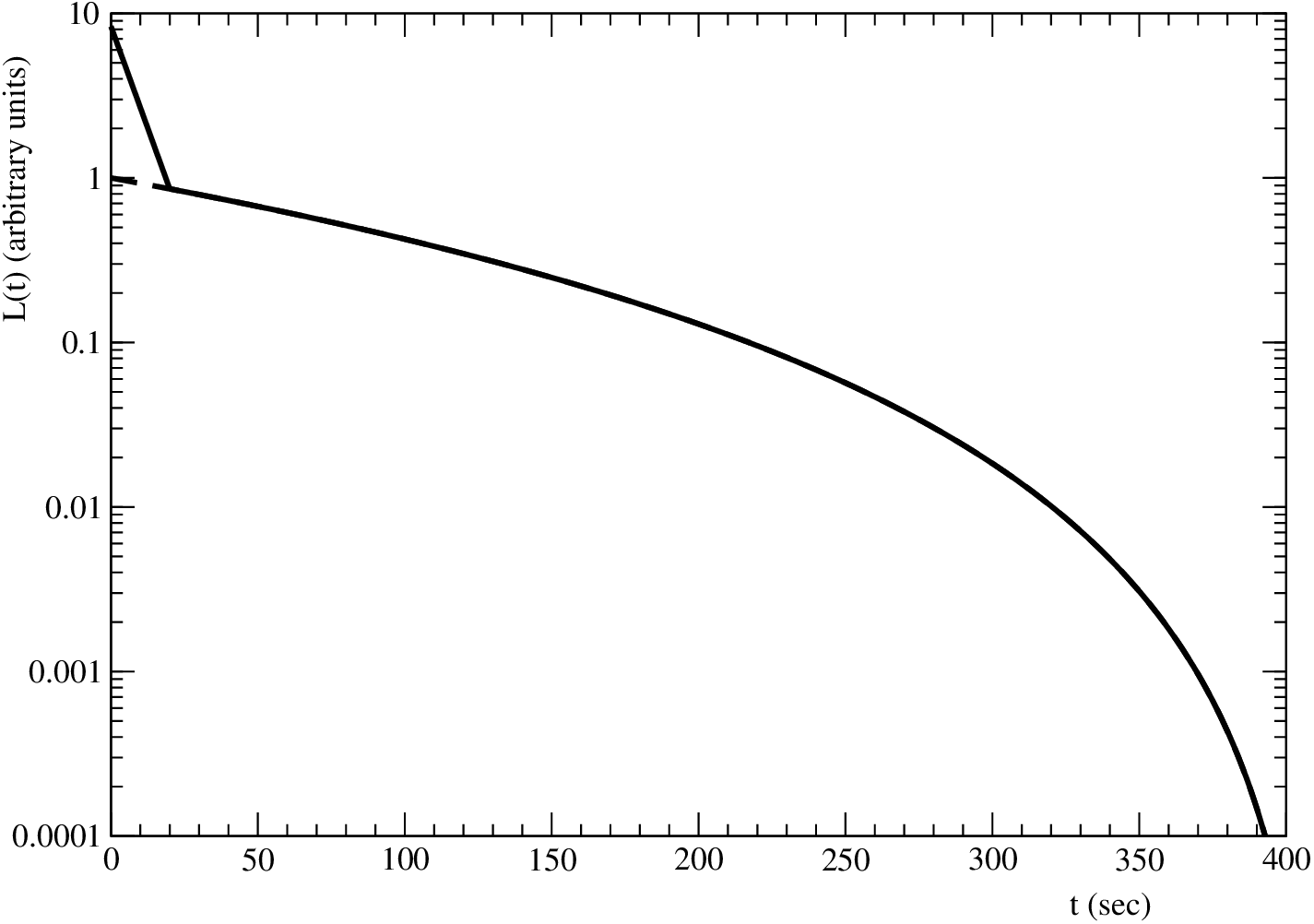}
\caption{\label{fig_5}
Light curve after the giant flare of 1998, August 27 from SGR 1900+14.  Dashed line is the light curve in the standard
magnetar model, Eq.~(\ref{5.7}) with parameters given in Eq.~(\ref{5.9}). Full line is our light curve Eq.~(\ref{5.6})
with parameters in  Eq.~(\ref{5.10})}
\end{figure}
\begin{figure}[ht]
   \centering
\includegraphics[width=0.9\textwidth,clip]{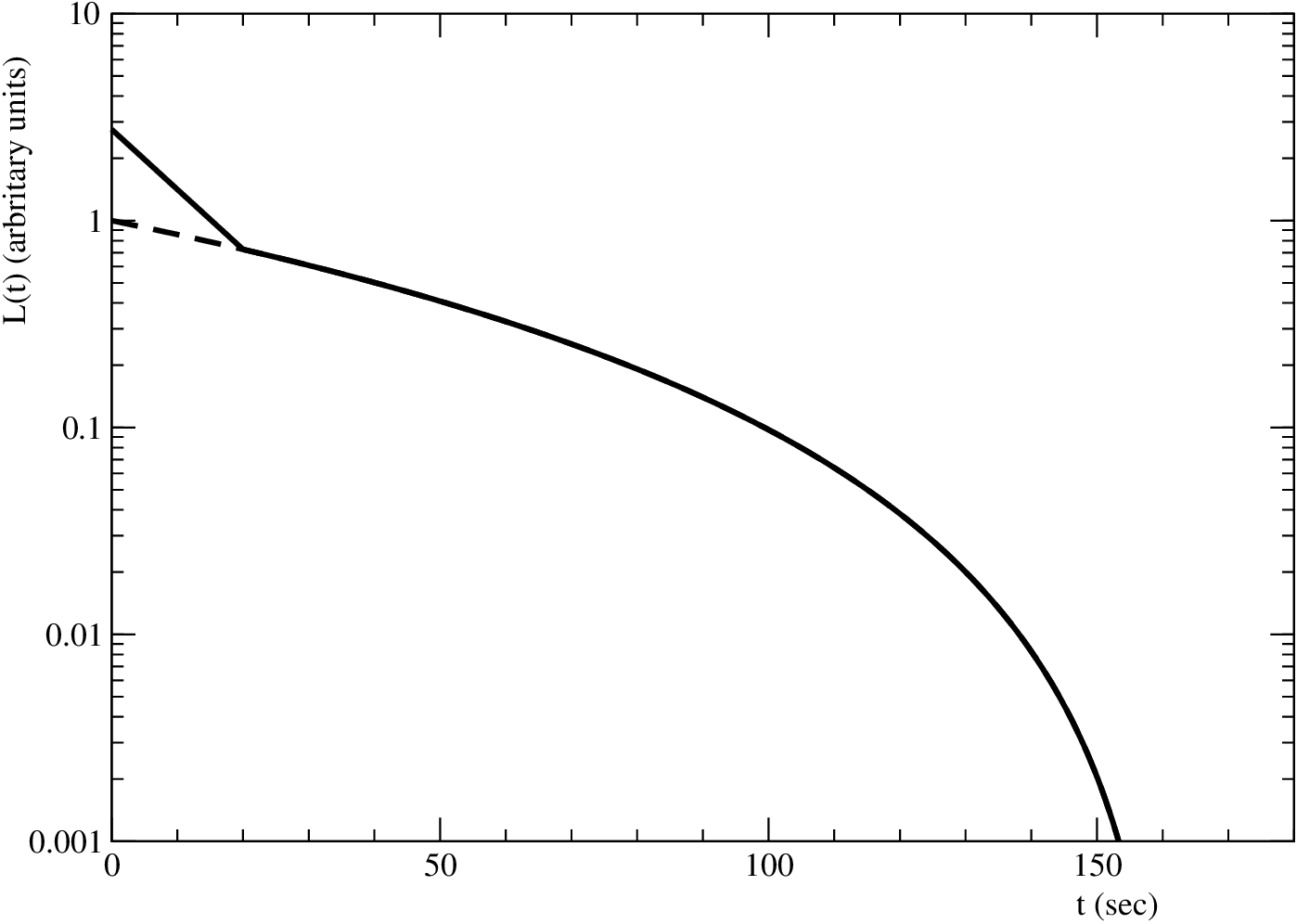}
\caption{\label{fig_6}
Light curve after the giant flare of 1979 March 5 from SGR 0526-66.  Dashed  line is the light curve in the standard
magnetar model, Eq.~(\ref{5.7}) with parameters given in Eq.~(\ref{5.11}). Full line is our light curve Eq.~(\ref{5.6})
with parameters in  Eq.~(\ref{5.12})}
\end{figure}
Feroci et al. (2001) performed a best fit of the light curve of the August 27 flare, background subtracted and binned to $
5 \, sec$, to Eq.~(\ref{5.7}) and found:
\begin{equation}
\label{5.9}
a \;  = \; 0.756 \; \pm \; 0.003 \; \; \; , \; \; \; t_{evap} \; = \;  414 \; sec  \; \; \;  .
\end{equation}
Indeed, from Fig.~2 in Feroci et al. (2001) one sees that the trapped fireball light curve account for the decay trend of
the experimental light curve and  matches the sudden final drop of the flux. However, it should stressed that the fitting
parameter $a$ in Eq.~(\ref{5.9}) does not satisfy the physical constraint Eq.~(\ref{5.8}).  Even more, any deviations from
spherical geometry or uniform temperature distribution lead to  parameters $a$ smaller than the upper bound $\frac{2}{3}$.
Moreover, the trapped fireball light curve underestimate by about an order of magnitude the measured flux during the first
stage of the outburst. We interpreted the different behavior of the flux during the initial phase of the outburst as a
clear indication of the linear regime described by our Eq.~(\ref{5.3}). As a matter of fact, we find that the measured
light curve could be better described by Eq.~(\ref{5.6}) with parameters (see Fig.~\ref{fig_5}):
\begin{equation}
\label{5.10}
\tau_0 \; = \; 8.80  \; sec \; \; , \; \; t_{break}  \; = \; 20 \; sec \; \; , \; \; \; \eta  \;  = \; 0.756 \; \; \; ,
 \; \; \;  t_{dis} \; = \;  414 \; sec  \; \; \; .
\end{equation}
The same criticisms apply to the fit within the standard magnetar model of the light curve after the giant flare of 1979
March 5 from SGR 0526-66. The trapped fireball light curve fit in Feroci et al. (2001) gives:
\begin{equation}
\label{5.11}
a \;  = \; 0.71 \; \pm \; 0.01 \; \; \; , \; \; \; t_{evap} \; = \; 163 \; \pm \; 5 \; sec  \; \; \;  .
\end{equation}
Again the parameter $a$ exceeds the bound Eq.~(\ref{5.8}), and the fit underestimates the flux during the first stage of
the outburst (see Fig.~14 in \cite{Feroci:2001}). Fitting our Eq.~(\ref{5.6}) to the measured flux reported in Fig.~14 in
Feroci et al. (2001), we estimate:
\begin{equation}
\label{5.12}
\tau_0 \; = \; 15  \; sec \; \; , \; \; t_{break}  \; = \; 20 \; sec \; \; , \; \; \; \eta  \;  = \; 0.71 \; \; \; ,
 \; \; \;  t_{dis} \; = \;  163 \; sec  \; \; \; .
\end{equation}
In Figs.~\ref{fig_5} and \ref{fig_6} we compare our light curves Eqs.~(\ref{5.6}), (\ref{5.10}) and (\ref{5.12}) with the
best fits performed in Feroci et al. (2001). Obviously, both light curves agree for $t > t_{break}$, while in the linear
regime $t < t_{break}$, where the trapped fireball light curves underestimate the flux, our light curves follow the
exponential decay and seem to be in closer agreement with observational data.

Several observations indicate that after a giant burst there are smaller and recurrent bursts. According to our theory
these small and recurrent bursts are the effect of several small glitches following the giant glitch. We may think about
these small bursts like the seismic activity following a giant earthquake. These seismic glitches are characterized by
light curves very different from the giant burst light curves. In the standard magnetar model these light curves are
accounted for with an approximate $t^{-0.7}$ decay (\cite{Lyubarsky:2002}). Within our theory there is a natural way to
describe the seismic burst activity.  Indeed, during these seismic bursts, that we shall call settling bursts, there is an
almost continuous injection of energy into the magnetosphere which tends to sustain an almost constant luminosity. This
corresponds to an effective $\kappa$ in Eq.~(\ref{5.1}) which decreases smoothly with time. The simplest choice is:
\begin{equation}
\label{5.13}
\kappa(t) \; = \; \frac{\kappa_0}{1 + \kappa_1 t}  \; \; \; .
\end{equation}
Inserting into  Eq.~(\ref{5.1}) and integrating, we get:
\begin{equation}
\label{5.14}
 E(t)  \; = \; \left [ E_0^{1-\eta} \; - (1 - \eta) \;  \frac{\kappa_0}{\kappa_1} \; \ln (1 + \kappa_1 t)
  \right ]^{\frac{\eta}{1-\eta}} \; \; \; .
\end{equation}
So that the luminosity is:
\begin{equation}
\label{5.15}
 L(t)  \; = \; \frac{L_0}{(1 + \kappa_1 t)^{\eta}} \; \left [1  \; -  (1 - \eta) \;  \frac{\kappa_0}{\kappa_1 E_0^{1-\eta}}
          \; \ln (1 + \kappa_1 t) \right ]^{\frac{\eta}{1-\eta}} \; \; \; .
\end{equation}
After defining the dissipation time:
\begin{equation}
\label{5.16}
 \ln (1 + \kappa_1 t_{dis}) \; =  \;  \frac{\kappa_1}{\kappa_0 }\; \frac{E_0^{1-\eta}}{1 - \eta} \; \; \; ,
\end{equation}
we rewrite Eq.~(\ref{5.15}) as
\begin{equation}
\label{5.17}
 L(t)  \; = \; \frac{L_0}{(1 + \kappa_1 t)^{\eta}} \; \left [ 1  \; -   \;  \frac{\ln (1 + \kappa_1 t)}{\ln (1 + \kappa_1
           t_{dis})}  \right ]^{\frac{\eta}{1-\eta}} \; \; \; .
\end{equation}
Note that the light curve Eq.~(\ref{5.17}) depends on two characteristic time constants $\frac{1}{\kappa_1}$ and
$t_{dis}$. We see that $\kappa_1 t_{dis}$, which is roughly the number of small bursts occurred in the given event, gives
an estimation of the seismic burst intensity. Moreover, since during the seismic bursts the injected energy is much
smaller than in giant bursts, we expect that fitting Eq.~(\ref{5.17}) to the observed light curves will result in values
of $\eta$ smaller than the typical values in giant bursts. In the following Sections we show that, indeed, our light
curves Eq.~(\ref{5.17}) are in good agreement with several observations.
\subsection{\normalsize{AXP 1E 2259+586}}
\label{2259}
On 2002, June 18 SGR-like bursts was recorded from AXP 1E 2259+586. Coincident with the burst activity were gross changes
in the pulsed flux, persistent flux, energy spectrum, pulse profile and spin down of the source (\cite{Woods:2004a}). As
discussed in previous Sections, these features are naturally accounted for within our theory. However, we believe that the
most remarkable and compelling evidence for our proposal comes from the observed coincidence of the burst activity with a
large glitch. Moreover, the time evolution of the unabsorbed flux from AXP 1E 2259+586 following the 2002 June outburst
reported in Woods et al. (2004) can be explained naturally within our theory. We could consider the June 18 SGR-like
bursts from AXP 1E 2259+586 the Rosetta Stone for our magnetar theory. \\
The temporal decay of the flux during the burst activity displays a rapid initial decay which lasted about $1 \; days$,
followed by a more mild decay during the year following the onset of the burst activity. Indeed, Woods et al. (2004)
splitted the data into two segments, and fit each independently to a power law:
\begin{equation}
\label{5.19}
\begin{split}
F(t) \; &  \; \;  \thicksim \; \;  t^{\alpha_1}  \;   \; , \; \alpha_1 \; = \; - \; 4.8 \; \pm \; 0.5 \; \; \; \; \;
                  \; , \; \;  t  \;  \lesssim 1 \; \; days \;  \; \;  ,  \\
F(t) \; &   \; \;  \thicksim \; \;  t^{\alpha_2}  \;   \; , \; \alpha_2 \; = \; - \; 0.22 \; \pm \; 0.01 \; \; \, ,
            \; \; t  \;   \gtrsim 1 \; \; days \;  \; \;    .
\end{split}
\end{equation}
\begin{figure}[ht]
   \centering
\includegraphics[width=0.9\textwidth,clip]{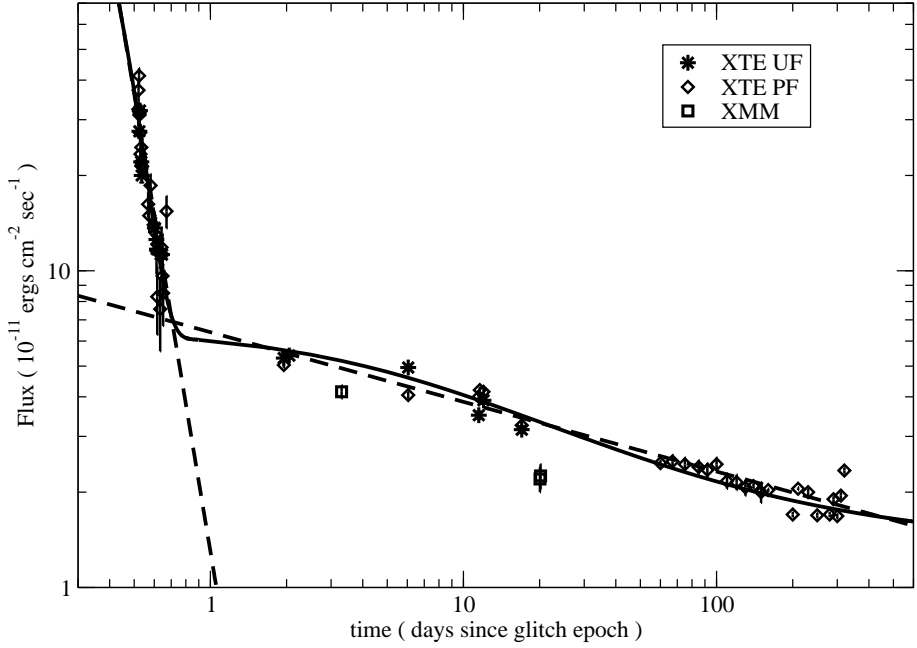}
\caption{\label{fig_7}
The time evolution of the unabsorbed flux from AXP 1E 2259+586 following the 2002 June outburst. Data have been extracted
from Fig.~13 in Woods et al. (2004).  Dashed  lines are the phenomelogical power law fits, Eq.~(\ref{5.19}). Full line is
our light curve Eq.~(\ref{5.21}), with parameters in Eq.~(\ref{5.24}).}
\end{figure}
It is evident from Eq.~(\ref{5.19}) that the standard magnetar model is completely unable to reproduce the phenomelogical
power law fit. On the other hand, even the phenomenological parametrization cannot account for the time evolution of the
flux. Indeed, if we assume the power law Eq.~(\ref{5.19}) for the decay of the flux, then we cannot explain why and how
the source returns in its quiescent state with quiescent flux (\cite{Woods:2004a}):
\begin{equation}
\label{5.20}
F_Q  \;  \;  \simeq  \; \;  1.53 \;  10^{-11} \; \frac{ergs}{ cm^{2} \, sec}  \; \; .
\end{equation}
Note that adding the quiescent flux to the power law decay does not resolve the problem, for in that case the fits worst
considerably. Our interpretation of the puzzling light curve displayed in Fig.\ref{fig_7} is that AXP 1E 2259+586 has
undergone a giant burst at the glitch epoch, and soon after the pulsar has entered into a intense seismic burst activity.
Accordingly, the flux can be written as:
\begin{equation}
\label{5.21}
 F(t) \; = \;  F_{GB}(t) \;  + \; F_{SB}(t)  \; +  \; F_Q \; \; \; \; ,
\end{equation}
where $F_Q$ is the quiescent flux, Eq.~(\ref{5.20}), $F_{GB}(t)$ is the giant burst contribution to the  flux given by
Eq.~(\ref{5.6}), and $F_{SB}(t)$ is the seismic  burst contribution  given by Eq.~(\ref{5.17}). Since during the first
stage of the outburst there are no available data, we may parameterize the giant burst contribution as:
\begin{equation}
\label{5.22}
F_{GB}(t) \; = \;  F_{GB}(0) \;  \left ( 1 - \frac{t}{t_{GB}} \right )^{\frac{\eta_{GB}}{1-\eta_{GB}}} \; \; ,
               \; \; \;  0 \; < \; t \; < \; t_{GB}  \; \;  ,
\end{equation}
while $F_{SB}(t)$ is given by:
\begin{equation}
\label{5.23}
 F_{SB}(t)  \; = \; \frac{F_{SB}(0)}{(1 + \kappa_1 t)^{\eta_{SB}}} \;
          \left [ 1  \; -   \;  \frac{\ln (1 + \kappa_1 t)}{\ln (1 + \kappa_1
           t_{SB})}  \right ]^{\frac{\eta_{SB}}{1-\eta_{SB}}}  \; \; ,
               \; \; \;  0 \; < \; t \; < \; t_{SB} \; \; \; ,
\end{equation}
where $t_{GB}$ and $t_{SB}$ are the dissipation time for giant and seismic bursts respectively. In Fig.~\ref{fig_7} we
display our light curve Eq.~(\ref{5.21}) with the following parameters:
\begin{equation}
\label{5.24}
\begin{split}
F_{GB}(0) \; &  \simeq \; 1.5 \;  10^{-8} \; \frac{ergs}{ cm^{2} \, sec} \; \; , \; \;   \eta_{GB} \; \simeq \;
             0.828 \; \; , \; \; t_{GB} \;  \simeq \;  0.91 \; days \;   \\
F_{SB}(0) \; &  \simeq \; 5.0 \;  10^{-11} \; \frac{ergs}{ cm^{2} \, sec} \;  , \; \eta_{SB} \; \simeq \; 0.45
                 \; ,  \; t_{SB} \;  \simeq \;  10^3 \; days \;  ,  \; \kappa_1  \; \simeq \; 0.20 \; days^{-1}  \; .
\end{split}
\end{equation}
A few comments are in order. First, the agreement with data is rather good. Second, our efficiency exponent $\eta_{GB}$ is
consistent with the values found in the giant bursts from  SGR 1900+14 and SGR 0526-66. On the other hand, quite
consistently, we have $ \eta_{SB} < \eta_{GB}$. Finally, we stress that from our interpretation of the light curve it
follows that the onset of the intense seismic burst activity ($\kappa_1 t_{SB} \sim 200$) did not allow a reliable
estimation of  $\frac{\delta \dot{\nu} }{\dot{\nu}}$, which we
predicted to be of order $10^{-2}$. \\
\begin{figure}[ht]
   \centering
\includegraphics[width=0.9\textwidth,clip]{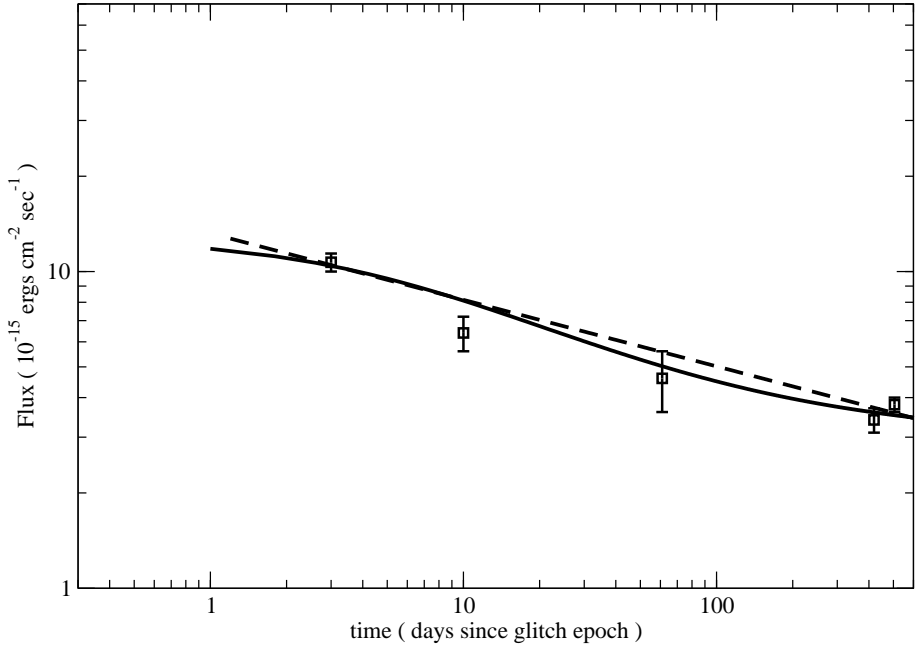}
\caption{\label{fig_8}
The time evolution of the unabsorbed IR flux from AXP 1E 2259+586 following the 2002 June outburst. Data have been
extracted from Fig.~1 in Tam et al. (2004). Dashed line is the phenomelogical power law fit $t^{-0.21 \pm 0.01}$. Full
line is our light curve Eq.~(\ref{5.25}).}
\end{figure}
Interestingly enough, following the 2002 June outburst it was detected an infrared flux change correlated with the $X$-ray
flux variability (\cite{Tam:2004}). Since the observations began three days after the 2002 June outburst, according to our
theory the infrared flux is parameterized as:
\begin{equation}
\label{5.25}
 F_{SB}^{IR}(t)  \; = \; \frac{F_{SB}^{IR}(0)}{(1 + \kappa_1 t)^{\eta_{SB}}} \;
          \left [ 1  \; -   \;  \frac{\ln (1 + \kappa_1 t)}{\ln (1 + \kappa_1
           t_{SB})}  \right ]^{\frac{\eta_{SB}}{1-\eta_{SB}}}  \; \; + \; \; F_{Q}^{IR} \; \; ,
               \; \; \;  0 \; < \; t \; < \; t_{SB} \; \; ,
\end{equation}
with the same parameters as in Eq.~(\ref{5.23}). Indeed, assuming $F_{SB}^{IR}(0) \simeq  9.5 \;  10^{-15}
\frac{ergs}{cm^{-2} sec^{-1}}$ and $F_{Q}^{IR} \simeq  3.3 \; 10^{-15} \frac{ergs}{cm^{-2} sec^{-1}}$, we found that our
light curve Eq.~(\ref{5.25}) is in remarkable good agreement with data (see Fig.~\ref{fig_8}). The strong correlation
between infrared and $X$-ray flux decays observed after the 2002 June outburst from AXP 1E 2259+586 strongly suggests a
physical link between the origin of both type of radiation.
\subsection{\normalsize{SGR 1900+14}}
\label{1900}
Soon after the 1998, August 27 giant burst, the soft gamma repeater SGR 1900+14  entered a remarkable phase of activity.
On August 29 an unusual burst from SGR 1900+14 was detected (\cite{Ibrahim:2001}) which lasted for a long time $\thicksim
10^{3} sec$. As discussed in Ibrahim et al. (2001), on observational grounds it can be ruled out extended afterglow tails
following ordinary bursts.
\begin{figure}[ht]
   \centering
\includegraphics[width=0.9\textwidth,clip]{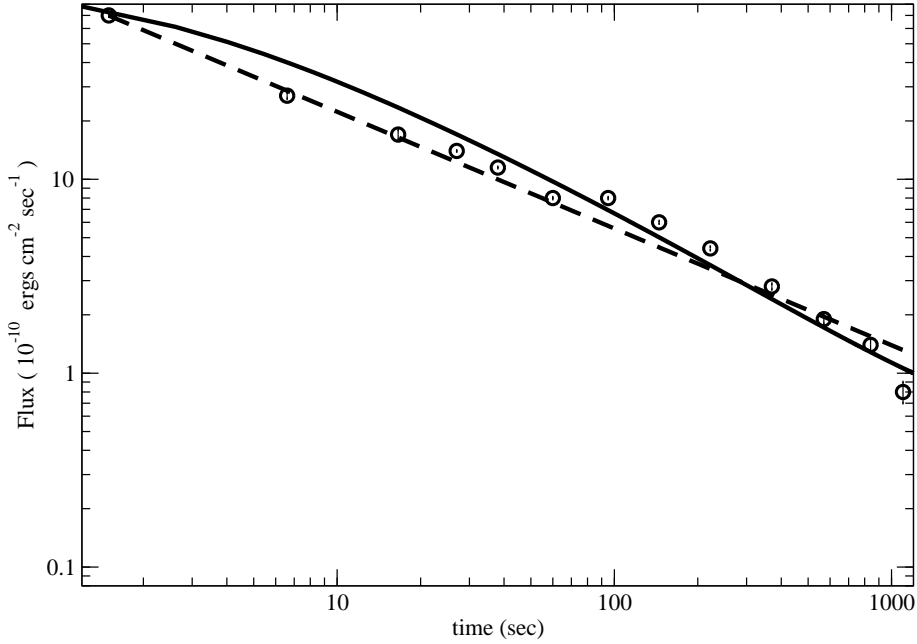}
\caption{\label{fig_9}
Flux evolution of the August 29 burst from SGR 1900+14. Data has been extracted from Fig.~4, panel (d), in Ibrahim et al.
(2001). Dashed line is the phenomenological power law fit Eq.~(\ref{5.26}); full line is our light curve Eqs.~(\ref{5.28})
and (\ref{5.29}).}
\end{figure}
\begin{figure}[ht]
   \centering
\includegraphics[width=0.9\textwidth,clip]{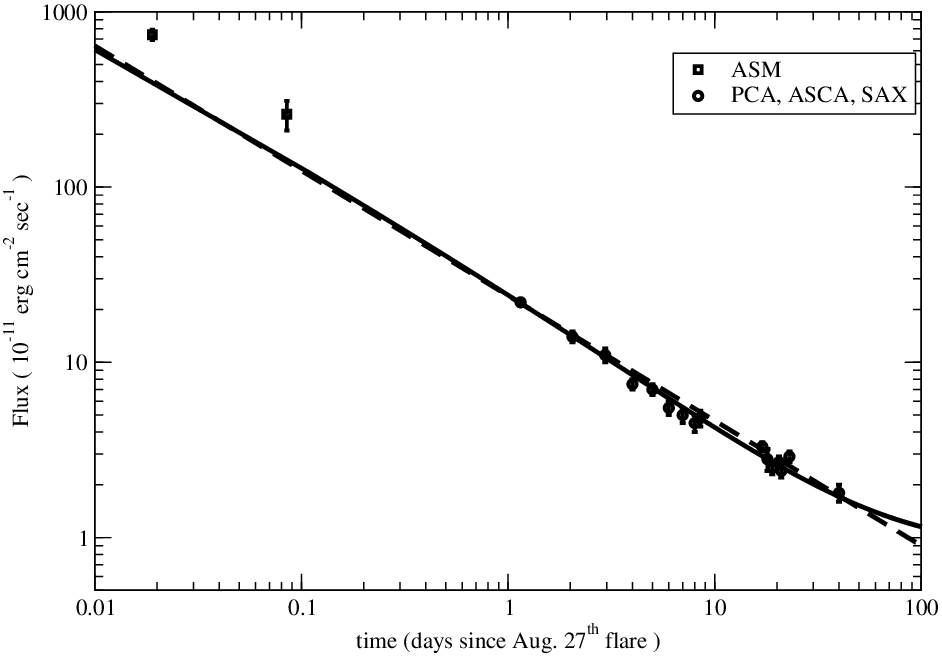}
\caption{\label{fig_10}
The time evolution of the unabsorbed flux from SGR 1900+14 following the 1998 August outburst. Data has been extracted
from Fig.~2 in Woods et al. (2001). Dashed line is the power law best fit $F(t) \thicksim t^{-(0.713 \pm 0.025)}$. Full
line is our light curve Eqs.~(\ref{5.28}) and (\ref{5.30}).}
\end{figure}
In Figure~\ref{fig_9} we display the flux decay after the August 29 burst. Data has been extracted from Ibrahim et al.
(2001). In Ibrahim et al. (2001) the temporal behavior of the flux decay has been parameterized as a power law (dashed
line in Fig.~\ref{fig_9}):
\begin{equation}
\label{5.26}
F(t)   \;  \;  = \; \; (89.16 \; \pm \; 1.34)  \; 10^{-10}  \; \frac{ergs}{ cm^{2} \, sec}  \;  t^{-(0.602 \pm 0.025)}\; .
\end{equation}
As already stressed, the phenomenological power law decay cannot explain the return of the source in its quiescent state
where the flux is (\cite{Hurley:1999a}):
\begin{equation}
\label{5.27}
F_Q  \;  \;  =  \; \;  0.96 \; \pm 0.07 \;  10^{-11} \; \frac{ergs}{ cm^{2} \, sec}  \; \; .
\end{equation}
On the other hand, we may easily account for the observed flux decay by our light curve:
\begin{equation}
\label{5.28}
 F(t)  \; = \; \frac{F(0)}{(1 + \kappa_1 t)^{\eta}} \;
          \left [ 1  \; -   \;  \frac{\ln (1 + \kappa_1 t)}{\ln (1 + \kappa_1
           t_{dis})}  \right ]^{\frac{\eta}{1-\eta}}  \; \; + \; \; F_Q \; \; ,
\end{equation}
where $F_Q$ is fixed by Eq.~(\ref{5.27}). Indeed, in Fig.~\ref{fig_9} we compare our light curve Eq.~(\ref{5.28}) with
observational data. The agreement is quite satisfying if we take:
\begin{equation}
\label{5.29}
F(0) \;  \simeq \; 1.05 \;  10^{-9} \; \frac{ergs}{ cm^{2} \, sec} \;  , \; \eta \; \simeq \; 0.5
                 \; ,  \; t_{dis} \;  \simeq \; 1.2 \; 10^3 \; sec \;  ,  \; \kappa_1  \; \simeq \; 0.50 \; sec^{-1}  \; .
\end{equation}

Woods et al. (2001) have analyzed a large set of $X$-ray observations of SGR 1900+14 in order to construct a more complete
flux history. They found that the flux level was more than an order of magnitude brighter than the level during
quiescence. This transient flux enhancement lasts about $40 \; days$ after the giant flare. Unlike Woods et al. (2001),
that argued that this enhancement was an artifact of the August 27 flare, we believe that the flux history can be
adequately described as seismic burst activity of the source. In Fig.~\ref{fig_10} we report the flux light curve
extracted from Fig.2 in Woods et al. (2001)  together with their power law best fit. Again we find the the flux history is
accounted for quite well by our light curve Eq.~(\ref{5.28}) with the following parameters:
\begin{equation}
\label{5.30}
F(0) \;  \simeq \; 4.8 \;  10^{-8} \; \frac{ergs}{ cm^{2} \, sec} \;  , \; \eta \; \simeq \; 0.55
                 \; ,  \; t_{dis} \;  \simeq \; 200 \; days \;  ,  \; \kappa_1  \; \simeq \;2 \; 10^3  \; days^{-1}  \; .
\end{equation}
The agreement between our light curve Eqs.~(\ref{5.28}) and (\ref{5.30}) with the power law best fit is striking. Moreover
we see that our curve deviates from the power law fit for $t > 60 \; days$ tending  to $F_Q$ at $t = t_{dis}$.
Woods et al. (2001) noted that extrapolating the fit to the August 27 $X$-ray light curve back toward the flare itself,
one finds that the expected flux level lies below the ASM flux measurements (squares in Fig.~\ref{fig_10}). Moreover,
these authors observed that the discrepancy reduces somewhat when one pushes forward the reference epoch to about 14
minutes after the onset of the flare. We believe that the discrepancy is due to a true physical effect, namely the
observed discrepancy from extrapolated light curve and ASM measurements is a clear indication that the surface luminosity
increases after the burst activity. In particular, soon after the  August 27 giant flare we have seen in
Sect.~\ref{bursts} that the surface temperature increases up to $\thicksim 61 \; KeV$ and the surface luminosity reaches
$\thicksim 10^{44} \; \frac{erg}{sec}$. Almost all the deposited energy is dissipated within the dissipation time of the
giant flare $\thicksim 400 \; sec$. Nevertheless, it is natural to expect a more gradual afterglow where a small fraction
of the energy deposited onto the star surface is gradually dissipated. As a matter of fact, we find that the observed
level of luminosity  $L_X \, \thicksim \, 10^{38} \; \frac{erg}{sec}$ (assuming a distance $d = 10 \; kpc$) at about $0.01
\; days$ since the August 27 giant flare, is consistent with the gradual afterglow scenario.\\
\begin{figure}[ht]
   \centering
\includegraphics[width=0.9\textwidth,clip]{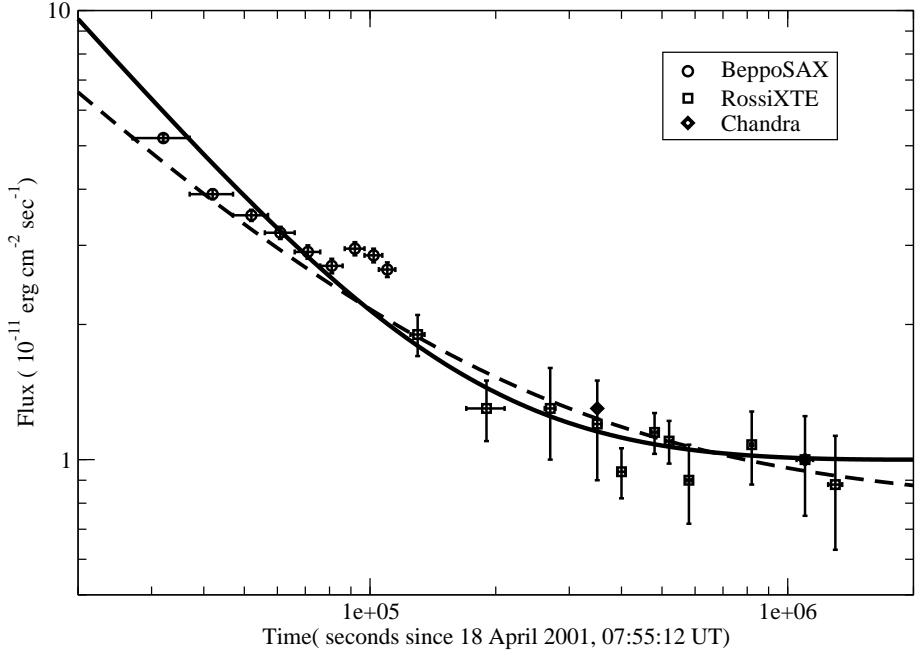}
\caption{\label{fig_11}
Temporal behavior of the $X$-ray flux from SGR 1900+14 in the aftermath of the 2001 April 18 flare. Data have been
extracted from Fig.~2 in Feroci et al. (2003). Dashed line is the power law best fit Eqs.~(\ref{5.31}) and (\ref{5.32}).
Full line is our light curve Eqs.~(\ref{5.28}) and (\ref{5.33}).}
\end{figure}
\begin{figure}[ht]
   \centering
\includegraphics[width=0.9\textwidth,clip]{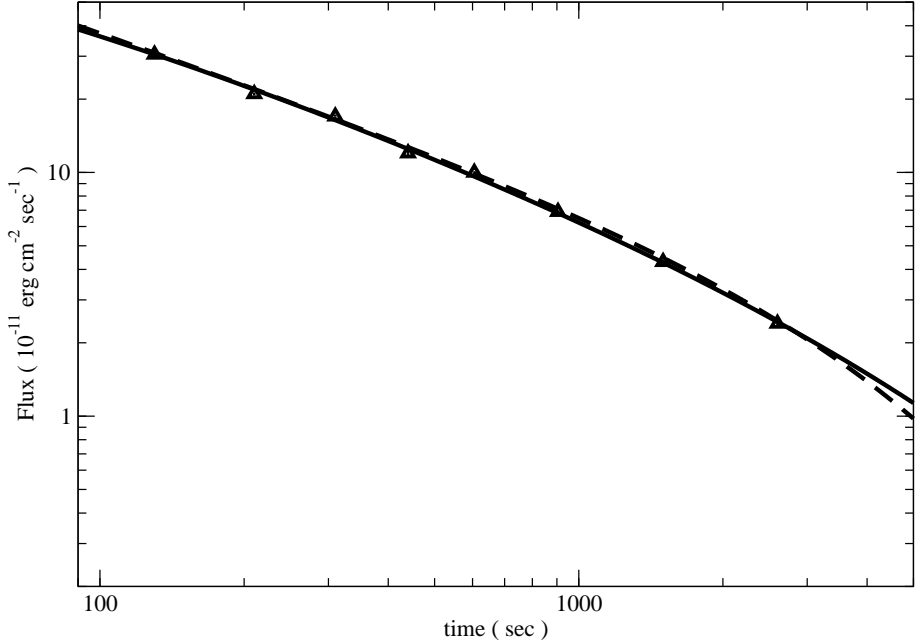}
\caption{\label{fig_12}
The temporal decay of the flux from the 2001 April 28 burst from SGR 1900+14 in the energy band $2 - 20 \; KeV$. The data
has been extracted from Fig.~5 in Lenters et al. (2003). Dashed line is the phenomenological best-fit power law times
exponential function adopted in Lenters et al. (2003) to describe data, Eqs.~(\ref{5.34}) and (\ref{5.35}). Full
continuous line is our light curve Eqs.~(\ref{5.28}) and (\ref{5.36}).
 }
\end{figure}
\begin{figure}[ht]
   \centering
\includegraphics[width=0.9\textwidth,clip]{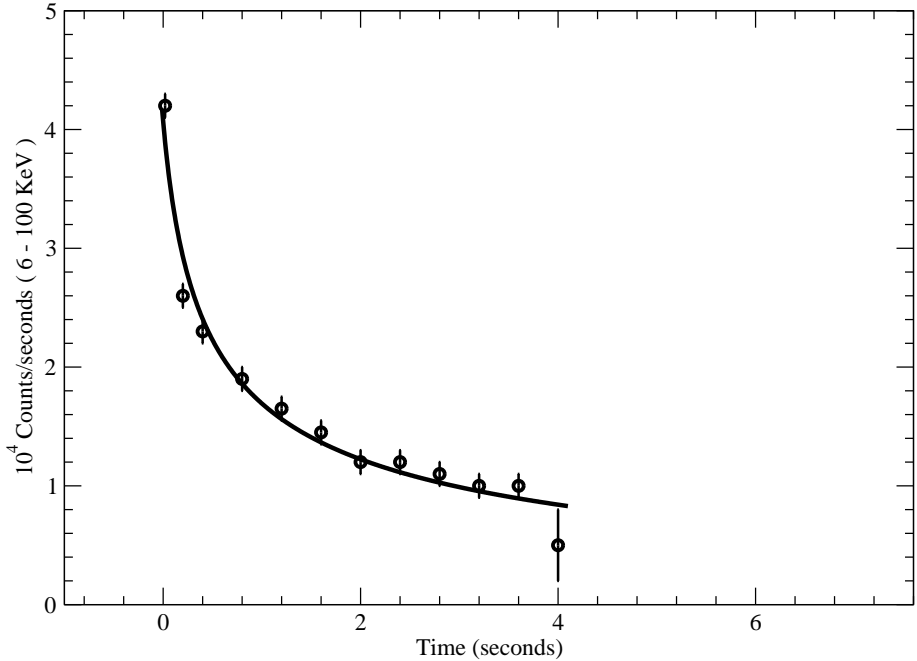}
\caption{\label{fig_14}
Time history of the 2001 July 2 burst from SGR 1900+14 in the energy band $7 - 100 \; KeV$ as observed by FREGATE. The
data has been extracted from Fig.~1 in Olive et al. (2004) by binning the light curve histogram. Full line is our light
curve Eqs.~(\ref{5.28}) and (\ref{5.39}).
}
\end{figure}
On 2001 April 18 the soft gamma ray repeater SGR 1900+14  emitted an intermediate burst. The light curve of this event did
not show any initial hard spike and was clearly spin-modulated. Moreover, the energetics appeared to be intermediate in
the $40-700 \, KeV$ range, with a total emitted energy of about $1.9 \; 10^{42} \; ergs$ (\cite{Feroci:2003}). In
Fig.~\ref{fig_11} we report the temporal behavior of the $X$-ray ($2 - 10 \; KeV$) flux from SGR 1900+14 in the aftermath
of the 2001 April 18 flare. Data has been extracted from Fig.~2 in Feroci et al. (2003).  Feroci et al. (2003) attempted a
simple power law function to the flux data:
\begin{equation}
\label{5.31}
 F(t)  \; \thicksim \; t^{-\alpha} \; \; + \; \; K \; \; ,
\end{equation}
where the constant $K$ should take care of the quiescent luminosity. Indeed,  Feroci et al. (2003), fitting
Eq.~(\ref{5.31}) to the data, found:
\begin{equation}
\label{5.32}
\alpha \; = 0.89(6) \; \; \; , \; \;  K \; = \; 0.78(5) \;  10^{-11} \; \; \frac{ergs}{ cm^{2} \, sec} \;  \; .
\end{equation}
As it is evident from Fig.~\ref{fig_11}, the power law  globally fits the data quite nicely. However, the reduced $\chi^2$
turns out to be in excess to $3$, mainly due to the bump occurring in the light curve at $t \thicksim 10^5 \, sec$
(\cite{Feroci:2003}). Indeed, after excluding the bump they get  a good fit with $\chi^2/dof \simeq 1$ without an
appreciable variation of the fit parameters (\cite{Feroci:2003}). However, there is still a problem with the
phenomelogical power law decay of the flux. As a matter of fact, Eq.~(\ref{5.32}) shows that the power law fit
underestimates the quiescent luminosity. In our opinion this confirms that the phenomenological power law decay of the
flux is not adeguate to describe the time variation of the flux. On the other hand, we find that our light curve
Eq.~(\ref{5.28}), with quiescent luminosity fixed to the observed value Eq.~(\ref{5.27}), furnishes a rather good
description of the flux decay once the parameters are given by:
\begin{equation}
\label{5.33}
F(0) \;  \simeq \; 2.6 \;  10^{-7} \; \frac{ergs}{ cm^{2} \, sec} \;  , \; \eta \; \simeq \; 0.68
                 \; ,  \; t_{dis} \;  \simeq \; 3 \; 10^6 \; sec \;  ,  \; \kappa_1  \; \simeq \; 0.25 \; sec^{-1}  \; .
\end{equation}
Within our interpretation, Eq.~(\ref{5.33}) shows that the flux decay in the aftermath of the April 18 flare is
characterized by a very large seismic burst activity ($\kappa_1 t_{dis} \thicksim 10^6$), which lasts for about $10^6 \;
sec$. So that the bump in the flux at  $t \thicksim 10^5 \, sec$ is naturally explained as  fluctuations in the intensity
of the seismic bursts.

Lenters et al. (2003) reported the spectral evolution and temporal decay of the $X$-ray tail of a burst from
 SGR 1900+14 recorded on 2001 April 28, 10 days after the intense April 18 event. In Fig.~\ref{fig_12} we display
the temporal decay of the flux from the 2001 April 28 burst in the energy band $2 - 20 \; KeV$. The data has been
extracted from Fig.~5 in  Lenters et al. These authors attempted several functional forms to fit the decay of the flux.
They reported that the decay was equally well fitted by either a power law times exponential or a broken power law. We
stress that both fits are phenomenological parametrization of the observational data, and that both fits are unable to
recover the quiescent flux. For definitiveness, we shall compare our light curve with the power law times exponential fit:
\begin{equation}
\label{5.34}
 F(t)  \; \thicksim \; t^{-\alpha} \; \exp(-\frac{t}{\tau}) \; \; \; \; ,
\end{equation}
The best fit to the temporal decay of the flux from the 2001 April 28 burst in the energy band $2 - 20 \; KeV$ gives
(\cite{Lenters:2003}):
\begin{equation}
\label{5.35}
\alpha \; = 0.68 \; \pm \; 0.04 \; \; \; \; , \; \; \; \; \tau \; = \; 5 \; \pm \; 1 \; 10^3 \; sec \; \; \; \; .
\end{equation}
In Figure~\ref{fig_12} we compare the phenomenological best fit Eqs.~(\ref{5.34}) and (\ref{5.35}) with our light curve
Eq.~(\ref{5.28}), where the quiescent luminosity is fixed to the observed value Eq.~(\ref{5.27}), and the parameters are
given by:
\begin{equation}
\label{5.36}
F(0) \;  \simeq \; 1.4 \;  10^{-9} \; \frac{ergs}{ cm^{2} \, sec} \;  , \; \eta \; \simeq \; 0.5
                 \; ,  \; t_{dis} \;  \simeq \; 5.5 \; 10^3 \; sec \;  ,  \; \kappa_1  \; \simeq \; 0.06 \; sec^{-1}  \; .
\end{equation}
Again, we see that our light curve gives a quite satisfying description of the flux decay. \\
Let us consider, finally, the light curve for the intermediate burst from SGR 1900+14 occurred on 2001 July 2
(\cite{Olive:2004}). In Figure~\ref{fig_14} we display the time decay of the flux after the July 2 burst. The data have
been extracted from Fig.~1 in Olive et al. (2004) by binning the light curve histogram. The displayed errors are our
estimate, so that the data should be considered as purely indicative of the decay of the flux. We find that Fig.~1 in
 Olive et al. (2004) is very suggestive, for it seems to indicate that the burst results from several small bursts, i.e.
according to our theory the burst is a seismic burst. As a consequence we try the fit with our light curve
Eq.~(\ref{5.28}). In this case the quiescent flux has been fixed to $F_Q \simeq 0$, for the observational data has been
taken in the energy range $7 - 100 \; KeV$ where the quiescent flux is very small. Attempting the fit to the data we find:
\begin{equation}
\label{5.39}
F(0) \;  \simeq \; 4.07 \;  10^{4} \; \frac{counts}{sec} \;  , \; \eta \; \simeq \; 0.36
                 \; ,  \; t_{dis} \;  \simeq \; 40 \;  sec \;  ,  \; \kappa_1  \; \simeq \; 5.0 \; sec^{-1}  \; .
\end{equation}
The resulting light curve is displayed in Fig.~\ref{fig_14}. The peculiarity of this burst resides in the fact that the
burst activity terminates suddenly at $t \simeq 4 \; sec$ well before the natural end at $t_{dis}  \simeq  40 \; sec $.
\subsection{\normalsize{SGR 1627-41}}
\label{1627}
\begin{figure}[ht]
   \centering
\includegraphics[width=0.9\textwidth,clip]{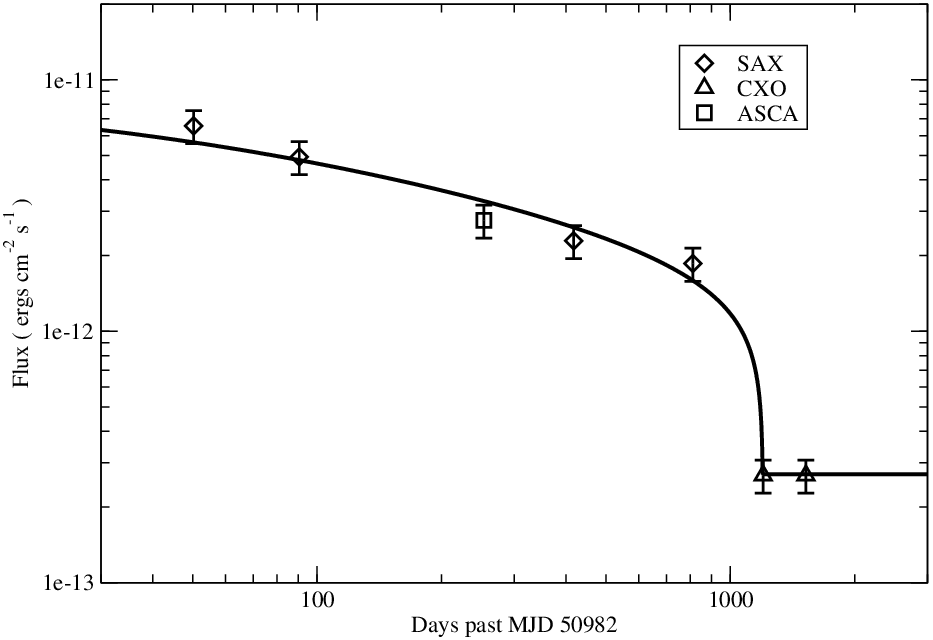}
\caption{\label{fig_15}
Time decay of the flux from SGR 1627-41. The data has been taken from Table~1 in Kouveliotou et al. (2003). Full line is
our best fitted light curve Eqs.~(\ref{5.28}), (\ref{5.40}) and (\ref{5.41}).
 }
\end{figure}
SGR 1627-41 was discovered with the Burst And Transient Source Experiment (BATSE) on the Compton Gamma-Ray Observatory
(CGRO) in June 1998 (\cite{Kouveliotou:1998,Woods:1999a}) when it emitted over 100 bursts within an interval of 6 weeks.
 Kouveliotou et al. (2003) presented the results of the monitoring of the flux decay of the $X$-ray
counterpart of SGR 1627-41 spanning an interval of roughly five years. Moreover, these authors attempted to understand
within the standard magnetar model the three year monotonic decline of SGR 1627-41 as cooling after a single deep crustal
heating event coinciding with the burst activity in 1998. They assumed an initial energy injection to the crust of the
order of $10^{44} \; ergs$. However, it must be pointed out that this assumption is highly unrealistic, for the estimate
of the total energy released in bursts during the activation of SGR 1627-41 ranges between $ 4 \; 10^{42} - 2 \; 10^{43}
\; ergs$. In addition, since gamma rays was not detected, they assumed that the conversion efficiency of the total energy
released during the activation into soft gamma rays were considerable less than $ 100 \; \%$. They also assumed that the
core temperature is low, i.e. the core cools via the direct URCA process. Notwithstanding these rather ad hoc assumptions,
Kouveliotou et al. (2003) was unable to explain the March 2003 data point, which clearly showed that the flux did not
decay further (see Fig.~\ref{fig_15}). In other words, the levelling of the flux during the third year followed by its
sharp decline is a feature that is challenging the standard magnetar model based on neutron stars, and that begs for an
explanation within that model. On the other hand, we now show that the peculiar SGR 1627-41 light curve find a natural
interpretation within our theory. In Fig.~\ref{fig_15} we display the time decay of the flux. The data has been taken from
Table~1 in  Kouveliotou et al. (2003). In this case we are able to  best fit our light curve Eq.~(\ref{5.28}) to available
data. Since the number of observations is rather low, to get a sensible fit we have fixed the dissipation time to $1200 \;
days$ and the quiescent luminosity to the levelling value at $t \gtrsim 1200 \; days$:
\begin{equation}
\label{5.40}
F_Q \;  \simeq \; 2.7 \;  10^{-13} \; \frac{ergs}{ cm^{2} \, sec} \; \; \; , \; \;  \; t_{dis} \;  \simeq \; 1200 \; days
          \;  \; .
\end{equation}
The best fit of our light curve to data gives:
\begin{equation}
\label{5.41}
F(0) \;  = \; 0.83(11) \;  10^{-11} \; \frac{ergs}{ cm^{2} \, sec} \;  , \; \eta \; = \; 0.25(8)
                 \; ,   \; \kappa_1  \; = \; 0.04(1) \; days^{-1}  \; .
\end{equation}
with a reduced $\chi^2 \, \thicksim \, 1$. From Fig.~\ref{fig_15}, where we compare our best fitted light curve with data,
we see that our theory allow a quite satisfying description of the three year monotonic decline of the flux.
\subsection{\normalsize{SGR 1806-20}}
\label{1806}
SGR 1806-20 entered an active phase in 2003, culminating in a gigantic flare on 2004 December 27, with energy greatly
exceeding that of all previous events. Figure~3 in Hurley et al. (2005)  displays the time history of the flux averaged
over the rotation period of the pulsar soon after the giant flare. These authors fitted the light curve within the
standard magnetar model based on the evaporation of a fireball formed after the giant burst and trapped onto the stellar
surface, Eq.~(\ref{5.7}). The fit of the rotation smoothed curve to the fireball function gives (\cite{Hurley:2005}):
\begin{equation}
\label{5.42}
 a  \;  = \; 0.606 \;  \pm \; 0.003 \; \; \; ,
 \; \; \;  t_{evap} \; = \;  382 \; \pm \; 3 \; \; sec  \; \; \; .
\end{equation}
\begin{figure}[th]
   \centering
\includegraphics[width=0.9\textwidth,clip]{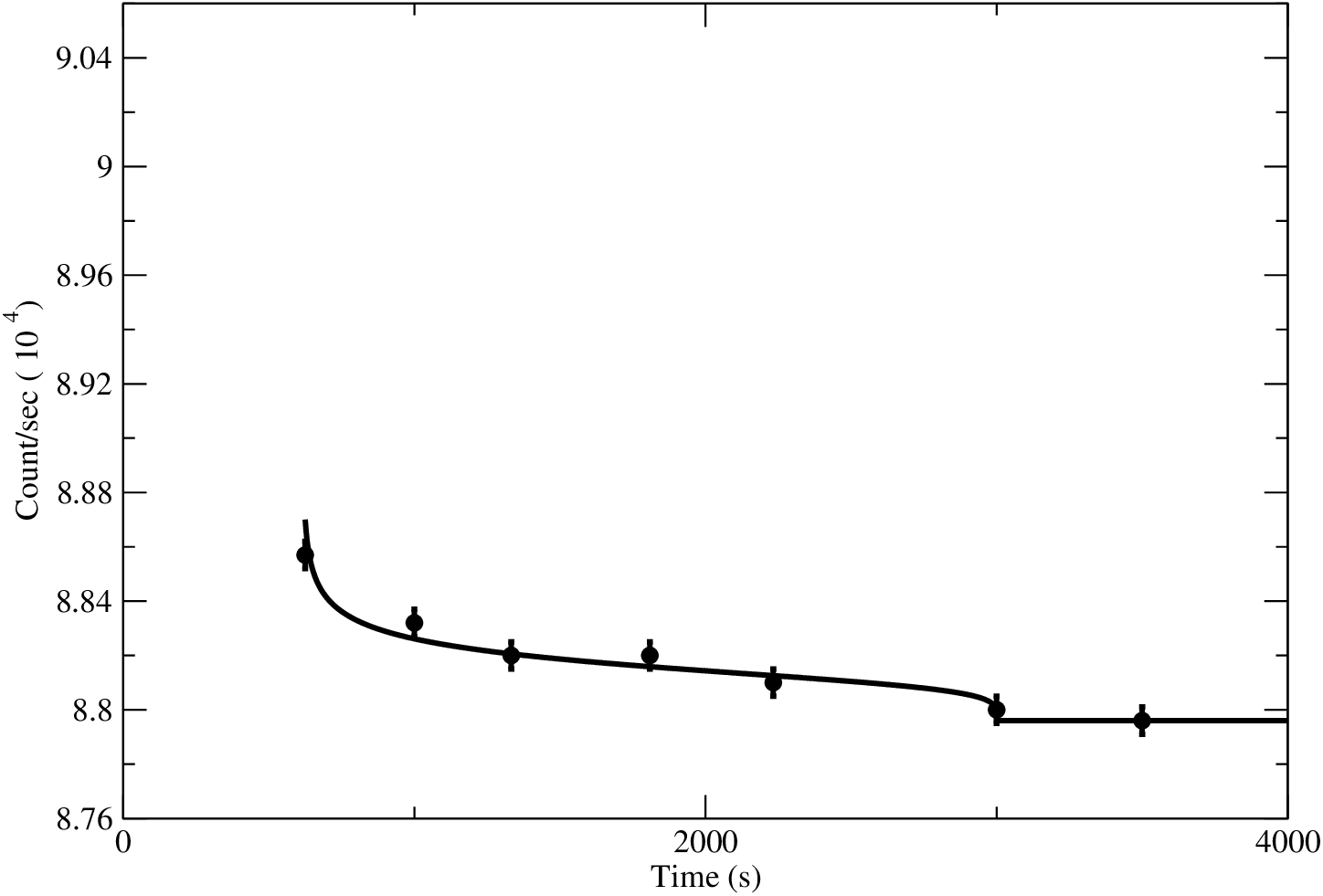}
\caption{\label{fig_16}
Time history of the second component after the 2004 December 27 giant burst from SGR 1806-20. The data has been extracted
from Fig.~5 in  Mereghetti et al. (2005b) by binning the light curve histogram. The continuous curve is our light curve
Eqs.~(\ref{5.28}), (\ref{5.43}) and (\ref{5.44}). }
\end{figure}
However, from Figure~3 in  Hurley et al. (2005) it is evident that fireball light curve underestimate the luminosity for
$t \lesssim 30 \; sec$. Thus we see that the light curve can be better accounted for by our light curve Eq.~(\ref{5.6})
with $t_{break} \simeq 30 \; sec$, quite close to the values found for the giant bursts from SGR 1900+14  and SGR 0526-66.
A more precise determination of the parameters of our light curve, however, must await for more precise data in the
initial phase of the afterglow. Instead, in the present Section we discuss the light curve of a second, separate component
after the giant burst reported in Mereghetti et al. (2005b). Indeed,  Mereghetti et al. (2005b) found evidence for a
separate component in the light curve starting at $t \thicksim 400 \; sec$ from the onset of the giant burst, forming a
peak at $t \thicksim 600 \; sec$ and ending at $t \thicksim 3000 \; sec$ (see Fig.~5 in \cite{Mereghetti:2005b}). As
already discussed, in our theory it is expected that there is an intense seismic burst activity following a giant burst.
In Figure~\ref{fig_16} we display the flux history starting from the giant flare. We show a few points of the second
component extracted from Fig.~5 in  Mereghetti et al. (2005b) by binning the light curve histogram. The displayed errors
are our estimate, so that the data should be considered as purely indicative of the decay of the flux. We fit the data
with our light curve Eqs.~(\ref{5.28}) assuming:
\begin{equation}
\label{5.43}
F_Q \;  \simeq \; 8.796 \;  10^{4} \; \frac{count}{sec} \; \;  , \; \; t_{dis} \; =  \; t_{end} \; -\; t_{start} \; \; ,
\; \; t_{start} \; \simeq \; 625 \; sec  \;  \; , \; \; t_{end} \; \simeq \; 3000 \; sec  \;  \; .
\end{equation}
The best fit of our light curve to data gives:
\begin{equation}
\label{5.44}
F(0) \;  = \; 0.074 \;  10^{4} \; \frac{count}{sec} \;  , \; \eta \; = \; 0.18
                 \; ,   \; \kappa_1  \; = \; 0.10 \; sec^{-1}  \; .
\end{equation}
In Fig.~\ref{fig_16} we compare our  light curve with data and  find that our theory allow a quite satisfying description
of the time history of the flux.
\section{\normalsize{CONCLUSIONS}}
\label{conclusion}
Let us summarize the main results of the present paper. We have discussed p-stars endowed with super strong dipolar
magnetic field. We found a well defined criterion to distinguish rotation powered pulsars from magnetic powered pulsars
(magnetars). We showed that glitches, that in our magnetars are triggered by magnetic dissipative effects in the inner
core, explain both the quiescent emission and bursts from soft gamma-ray repeaters and anomalous $X$-ray pulsars. In
particular, we were able to account for the braking glitch from SGR 1900+14 and the normal glitch from AXP 1E 2259+586
following a giant burst. We accounted for the observed anti correlation between hardness ratio and intensity. Within our
magnetar theory we were able to account quantitatively for light curves for both gamma-ray repeaters and anomalous $X$-ray
pulsars. In particular we explained the light curve after the June 18, 2002 giant burst from AXP 1E 2259+586. The ability
of our p-star theory to reach a satisfying  understanding of several observational features of soft gamma-ray repeaters
and anomalous $X$-ray pulsars supports our proposal for a revision of the standard paradigm
of relativistic astrophysics. \\
Let us conclude by briefly addressing the theoretical foundation of our theory. As a matter of fact, our proposal for
p-stars stems from recent numerical lattice results in QCD (\cite{cosmai:2003}), which suggested that the gauge system
gets deconfined in strong enough chromomagnetic field. This leads us to consider the new class of compact quark stars made
of almost massless deconfined up and down quarks immersed in a chromomagnetic field in $\beta$-equilibrium. Our previous
studies showed that these compact stars are more stable than neutron stars whatever the value of the chromomagnetic
condensate. This remarkable result indicates that the true ground state of QCD in strong enough gravitational field is not
realized by hadronic matter, but by p-matter. In other words, the final collapse of an evolved massive star leads
inevitably to the formation of a p-star.

\end{document}